\documentclass[journal]{IEEEtran}
\pdfoutput=1
\IEEEoverridecommandlockouts

\usepackage{acronym}
\usepackage{amsmath,amssymb}
\usepackage[english]{babel}
\usepackage{bbm}
\usepackage{bm}
\usepackage{caption}
\usepackage{cite}
\usepackage{enumerate}
\usepackage[T1]{fontenc}
\usepackage{graphicx}
\usepackage{hyperref}
\usepackage{cleveref}
\usepackage[utf8]{inputenc}
\usepackage{mathtools}
\usepackage{siunitx}
\usepackage{subcaption}
\usepackage{xcolor}
\usepackage{algorithm}
\usepackage{algorithmic}

\definecolor{myblue}{RGB}{31,119,180}
\definecolor{myorange}{RGB}{255,127,14}
\definecolor{mygreen}{RGB}{44,160,44}

\acrodef{CME}{chemical master equation}
\acrodef{PBS}{particle-based computer simulation}
\acrodef{NT}{neurotransmitter}
\acrodef{LNA}{linear noise approximation}
\acrodef{MC}{molecular communication}
\acrodef{DMC}{diffusive molecular communication}
\acrodef{LGIC}{ligand-gated ion channel}
\acrodef{PSP}{postsynaptic membrane potential}
\acrodef{AMPAR}{$\alpha$-amino-3-hydroxy-5-methyl-4-isoxazolepropionic acid receptor}
\acrodef{nAChR}{nicotinic acetylcholine receptor}
\acrodef{NMDAR}{$N$-methyl-\textsc{D}-aspartate receptor}
\acrodef{CoV}{coefficient of variation}

\DeclareMathOperator*{\argmax}{arg\,max}

\newcommand{\kd}{\ensuremath{\kappa_\mathrm{d}}}
\newcommand{\ke}{\ensuremath{\kappa_\mathrm{e}}}
\newcommand{\ka}{\ensuremath{\kappa_\mathrm{a}}}
\renewcommand{\P}{\ensuremath{P}}
\newcommand{\dt}{\ensuremath{\mathrm{d}t}}
\newcommand{\dtau}{\ensuremath{\mathrm{d}\tau}}
\newcommand{\kadt}{\ensuremath{\tilde{\kappa}_{\mathrm{a}_0}}}
\newcommand{\pibd}{\ensuremath{\boldsymbol{\pi}}}
\newcommand{\Abd}{\ensuremath{\boldsymbol{A}}}
\newcommand{\Qbd}{\ensuremath{\boldsymbol{Q}}}
\newcommand{\Dbd}{\ensuremath{\boldsymbol{D}}}
\newcommand{\Nmax}{\ensuremath{N_\textrm{max}}}
\newcommand{\Nmin}{\ensuremath{N_\textrm{min}}}
\newcommand{\Omax}{\ensuremath{O_\textrm{max}}}
\newcommand{\Omin}{\ensuremath{O_\textrm{min}}}
\newcommand{\Isc}{\ensuremath{I_\textrm{sc}}}
\newcommand{\gsc}{\ensuremath{\gamma_\textrm{sc}}}
\newcommand{\Vm}{\ensuremath{V_\textrm{m}}}
\newcommand{\Vmt}{\ensuremath{\tilde{V}_\textrm{m}}}
\newcommand{\Er}{\ensuremath{E_\textrm{r}}}
\newcommand{\Isyn}{\ensuremath{I_\textrm{syn}}}
\newcommand{\IL}{\ensuremath{I_\textrm{L}}}
\newcommand{\gL}{\ensuremath{g_\textrm{L}}}
\newcommand{\EL}{\ensuremath{E_\textrm{L}}}
\newcommand{\Cm}{\ensuremath{C_\textrm{m}}}
\newcommand{\Cov}{\ensuremath{\textrm{Cov}}}
\newcommand{\Var}{\ensuremath{\textrm{Var}}}
\newcommand{\Exp}[1]{\ensuremath{\mathbb{E}\left[#1\right]}}
\newcommand{\KOO}{\ensuremath{K_{OO}}}
\renewcommand{\S}{\ensuremath{\mathfrak{S}}}
\newcommand{\A}{\ensuremath{\textrm{A}}}
\newcommand{\DKL}{\ensuremath{\textrm{D}_\textrm{KL}}}

\newtheorem{theorem}{Theorem}

\makeatletter
\long\def\@makecaption#1#2{\ifx\@captype\@IEEEtablestring%
    \footnotesize\begin{center}{\normalfont\footnotesize #1}\\
        {\normalfont\footnotesize\scshape #2}\end{center}%
    \@IEEEtablecaptionsepspace
    \else
    \@IEEEfigurecaptionsepspace
    \setbox\@tempboxa\hbox{\normalfont\footnotesize {#1.}~~ #2}%
    \ifdim \wd\@tempboxa >\hsize%
    \setbox\@tempboxa\hbox{\normalfont\footnotesize {#1.}~~ }%
    \parbox[t]{\hsize}{\normalfont\footnotesize \noindent\unhbox\@tempboxa#2}%
    \else
    \hbox to\hsize{\normalfont\footnotesize\hfil\box\@tempboxa\hfil}\fi\fi}
\makeatother

\begin{document}
    
\nocite{lotter21a}
    
\title{Molecular Noise In Synaptic Communication
    \thanks{This paper has been presented in part at the IEEE Wireless Communications and Networking Conference (WCNC), 2022 \cite{lotter21a}. This work was supported in part by the German Research Foundation (DFG) under grant SCHO 831/9-1.}
}

\author{\IEEEauthorblockN{Sebastian Lotter, Maximilian Sch\"afer, and Robert Schober\\} 
    \IEEEauthorblockA{Friedrich-Alexander University Erlangen-N\"urnberg, Germany}
}

\maketitle

\begin{abstract}
    In synaptic \ac{MC}, the activation of postsynaptic receptors by \acp{NT} is governed by a stochastic reaction-diffusion process. %
    This randomness of synaptic \ac{MC} contributes to the randomness of the electrochemical downstream signal in the postsynaptic cell, called \ac{PSP}.
    Since the randomness of the \ac{PSP} is relevant for neural computation and learning, characterizing the statistics of the \ac{PSP} is critical. 
    However, the statistical characterization of the synaptic reaction-diffusion process is difficult because the reversible bi-molecular reaction of \acp{NT} with receptors renders the system nonlinear.
    Consequently, there is currently no model available which characterizes the impact of the statistics of postsynaptic receptor activation on the \ac{PSP}.
    In this work, we propose a novel statistical model for the synaptic reaction-diffusion process in terms of the \ac{CME}.
    We further propose a novel numerical method which allows to compute the \ac{CME} efficiently and we use this method to characterize the statistics of the \ac{PSP}.
    Finally, we present results from stochastic particle-based computer simulations which validate the proposed models.
    We show that the biophysical parameters governing synaptic transmission shape the autocovariance of the receptor activation and, ultimately, the statistics of the \ac{PSP}.
    Our results suggest that the processing of the synaptic signal by the postsynaptic cell effectively mitigates synaptic noise while the statistical characteristics of the synaptic signal are preserved.
    The results presented in this paper contribute to a better understanding of the impact of the randomness of synaptic signal transmission on neuronal information processing.
\end{abstract}
\acresetall

\section{Introduction}
\label{sec:introduction}
\Ac{DMC} is a novel communication paradigm inspired by the exchange of information between biological entities by means of diffusing molecules\cite{nakano13}.
It is envisioned that synthetic \ac{DMC} will enable revolutionary applications in the field of intra-body nano-scale communications based on and for interfacing with natural \ac{MC} systems, such as the synaptic \ac{DMC} system\cite{veletic2019}.
Since the synaptic \ac{DMC} system enables complex processes such as learning and memory, understanding the underlying design principles is key to the development of synthetic neural applications such as neural prostheses and brain-machine interfaces\cite{veletic2019}.
However, despite considerable research efforts over the last decades (see \cite{pickel14} and references therein), our picture of synaptic communication is not complete, yet \cite{rusakov20}.

In synaptic \ac{DMC}, information is conveyed from a {\em presynaptic} cell to a {\em postsynaptic} cell by means of diffusing molecules called \acp{NT}.
\acp{NT} are released by exocytosis from the presynaptic cell, bind reversibly to transmembrane receptors at the postsynaptic cell, and may be degraded by enzymes while diffusing in the extracellular medium \cite{zucker14}, cf.~Fig.~\ref{fig:synapse}.
The activation of ionotropic receptors, i.e., \acp{LGIC}, by \acp{NT} leads to a local depolarization of the postsynaptic membrane which propagates to the soma of the postsynaptic cell as an input to the computations carried out by the postsynaptic cell \cite{byrne14}.
The diffusion of \acp{NT} inside the synaptic cleft as well as the degradation of \acp{NT} and the activation of postsynaptic receptors are random processes.
Consequently, the depolarization of the postsynaptic membrane, termed {\em \ac{PSP}}, is a random process, too.
One central open question regarding synaptic neural communication concerns the impact of the randomness of the \ac{PSP} on neural information transmission \cite{rusakov20}.
Indeed, various roles for the randomness of the \ac{PSP} in neural communication have been suggested \cite{stacey01,aitchison21}.
However, current computational models of synaptic communication are not able to explain the stochastic variability of the \ac{PSP} \cite{baker11}.
This paper provides a step towards filling this research gap by studying the impact of the randomness of the synaptic reaction-diffusion process on the statistics of the \ac{PSP}.
In this way, the statistical model proposed in this paper contributes to a complete statistical characterization of the \ac{PSP} which may ultimately reveal the role of noise in synaptic neural communication.

Synaptic \ac{DMC} has been studied in the \ac{MC} community with emphasis on different aspects, such as information theoretic limits \cite{veletic20}, the design of artificial synapses \cite{bilgin17}, and the long-term average signal decay \cite{oncu21}, see also literature overviews in \cite{veletic2019,lotter20}.
Mean-field models, i.e., deterministic models for the average activation of postsynaptic receptors valid in the large system limit, have been developed for synapses employing enzymatic degradation \cite{lotter21,oncu21} and other channel clearance mechanisms \cite{khan2017,lotter20,bilgin17}.
However, stochastic fluctuations in the activation of postsynaptic receptors have been considered only recently \cite{lotter21}.
Yet, the statistical model proposed in \cite{lotter21} does not account for the randomness of the enzymatic degradation of \acp{NT} and relies on the simplifying assumption that either \acp{NT} compete for receptors or receptors compete for \acp{NT}.
Hence, the scope and applicability of the model in \cite{lotter21} is limited to a specific range of parameter values.
Statistical models for ligand-binding receptors employed in the \ac{MC} literature outside synaptic communication assume statistical independence of the receptors \cite{kuscu18} or require the concentration of solute molecules to be independent of the molecule binding \cite{pierobon11a} (see \cite{egan21} for a recent survey on modeling techniques for stochastic reaction-diffusion systems employed in the \ac{MC} literature).
As already shown in \cite{lotter21}, these assumptions are not always justified.
The impact of the random propagation and reaction of \acp{NT} on the \ac{PSP} has, to the best of the authors' knowledge, not been considered in previous studies.

In this paper, we propose a novel statistical signal model for synaptic \ac{DMC} in terms of the \ac{CME}.
The proposed model characterizes the joint statistics of the activation of postsynaptic \acp{LGIC} and the enzymatic degradation process for the first time in the \ac{MC} literature.
Furthermore, in contrast to existing models, it does not rely on simplifying assumptions with respect to the statistical (in)dependence between receptors and/or \acp{NT} and allows for the computation of the non-stationary autocovariance of the \ac{LGIC} activation.
Since the \ac{CME} model in its original form is computationally intractable, a novel adaptive state reduction scheme is proposed which allows the efficient computation of the proposed model.
The proposed state reduction scheme exploits knowledge of the first-order statistics of the considered process and, in contrast to common approximation methods for the \ac{CME} found in the literature\cite{munsky18}, the approximation error is explicitly characterized and can, hence, be controlled.
Using the proposed \ac{CME} model, the mean and the variance of the \ac{PSP} caused by the presynaptic release of \acp{NT} is characterized and an approximation of the instantaneous statistics of the \ac{PSP} in terms of the Gaussian distribution is proposed.
Since physical parameters of the synaptic \ac{DMC} system, such as the number of postsynaptic receptors and the chemical reaction constants, are reflected in the proposed model, the impact of these parameters on the statistics of the \ac{PSP} can be analyzed for the first time.
Finally, the results of the proposed model are compared to stochastic \acp{PBS} to validate the assumptions made to arrive at the proposed model and to verify the accuracy of the presented results.
In short, the main contributions of this paper can be summarized as follows:
\begin{enumerate}
    \item A \ac{CME}-based statistical model for the postsynaptic receptor activation and \ac{NT} degradation is proposed.
    \item The autocovariance function of the postsynaptic receptor activation is derived.
    \item The \ac{PSP} is characterized statistically in terms of the non-stationary receptor occupancy statistics.
    \item A novel, adaptive numerical algorithm to efficiently compute the \ac{CME} model is proposed.
    \item The derived results are validated by stochastic \acp{PBS} and used to study the impact of different synaptic configurations on the statistics of the \ac{PSP}.
\end{enumerate}
In summary, the proposed model allows for an accurate statistical characterization of the synaptic noise caused by \ac{NT} binding and degradation and its impact on the \ac{PSP}.
It hence provides a step forward towards a better understanding of the role of synaptic noise for neural information processing.

The \ac{CME} model for the postsynaptic receptor activation and the proposed adaptive state reduction scheme presented in this paper were introduced in part in \cite{lotter21a}.
However, the present paper extends the \ac{CME} model proposed in \cite{lotter21a} by a model for the autocovariance of the postsynaptic receptor activation.
Furthermore, while the \ac{PSP} was not considered in \cite{lotter21a}, it is approximated via a linear filter and characterized statistically in this paper.
In contrast to \cite{lotter21a}, the results presented in the present paper provide insight into the impact of the parameters of the synapse and the postsynaptic membrane, respectively, on the \ac{PSP}.
Hence, the results presented in this paper constitute a major extension of \cite{lotter21a}.

The remainder of this paper is organized as follows.
The system model is introduced in Section~\ref{sec:system_model}.
In Section~\ref{sec:psp}, the mean and the variance of the \ac{PSP} as a function of the stochastic activation of the postsynaptic receptors are derived and an approximation of the \ac{PSP} in terms of the Gaussian distribution is proposed.
In Section~\ref{sec:cme}, a state reduction scheme for the computation of the \ac{CME} introduced in Section~\ref{sec:system_model} is provided.
In Section~\ref{sec:numerical_results}, the proposed model is used to study the statistics of the \ac{PSP} for selected, biologically relevant parameter regimes and numerical results from \acp{PBS} are presented to validate the model.
Section~\ref{sec:conclusion} concludes the paper with a brief summary of the main findings and an outlook on future research directions.

\section{System Model}
\label{sec:system_model}
\subsection{Biological Background}\label{sec:system_model:biological_background}

\begin{figure*}[!tbp]
    \centering
    \includegraphics[width=.8\textwidth]{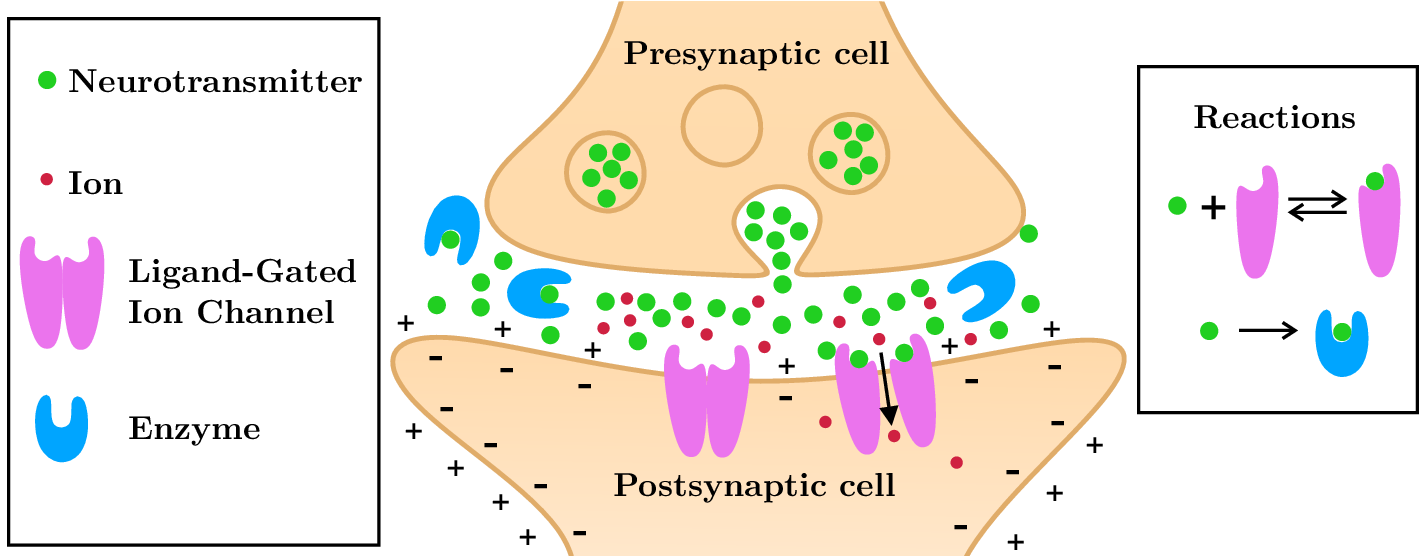}
    \caption{Chemical synapse. \acp{NT} (green) are released by exocytosis at the presynaptic cell and diffuse in the synaptic cleft. Solute \acp{NT} can bind reversibly to ionotropic receptors (pink) at the postsynaptic membrane and may be degraded by enzymes (blue). Activated receptors enable the flux of positively charged ions (red) through the polarized postsynaptic membrane. The two reactions considered for the statistical analysis in this paper are the reversible bimolecular reaction of \acp{NT} with postsynaptic receptors and the unimolecular degradation process modeling the degradation of \acp{NT} by enzymes.}
    \label{fig:synapse}
\end{figure*}

We consider two neural cells, a presynaptic cell and a postsynaptic cell, which communicate via a chemical synapse, cf.~Fig.~\ref{fig:synapse}.
The plasma membrane of the postsynaptic cell acts as a diffusion barrier for positively and negatively charged ions present in the extracellular environment and inside the postsynaptic cell, e.g., sodium ($\textrm{Na}^+$), potassium ($\textrm{K}^+$) (both positively charged), and chloride ($\textrm{Cl}^-$) (negatively charged).
Under resting conditions, i.e., in the absence of neurotransmission, membrane-bound ion pumps and ion channels acting independently of neurotransmission maintain an electrochemical gradient between the intracellular and the extracellular environment called the {\em resting potential} or {\em leakage potential} $\EL$ of the membrane \cite{byrne14}.
$\EL$ is negative at approximately $\SI{-60}{\milli\volt}$ to $\SI{-80}{\milli\volt}$ \cite{byrne14,kobayashi11} reflecting the relative abundance of negative charge in the intracellular space compared to the extracellular space under resting conditions.

During neurotransmission, \acp{NT} are released into the synaptic cleft by exocytosis of presynaptic vesicles.
After release, \acp{NT} propagate by Brownian motion and react with postsynaptic transmembrane receptors and degradative enzymes \cite{zucker14}, cf.~Fig.~\ref{fig:synapse}.
In excitatory synapses, the type of synapses considered in this paper, the activation of ionotropic receptors leads to the flux of positively charged ions from the extracellular space into the postsynaptic intracellular space \cite{byrne14}.
Since the postsynaptic membrane is negatively polarized under resting conditions, this transfer of charge causes a local depolarization of the membrane, called  \ac{PSP}.
Hence, the chemical signal carried by the \acp{NT} released at the presynaptic cell is converted into an electrical signal at the postsynaptic cell.
Since both the reactions and the diffusion of the \acp{NT} are random, the activation of postsynaptic receptors and, consequently, also the \ac{PSP} are random processes.

\subsection{A Deterministic Model for Synaptic DMC}

For synaptic \ac{DMC} systems satisfying the assumptions discussed in \cite[Sec.~II-A]{lotter21}, i.e., synapses that employ enzymatic degradation as channel clearance mechanism and are either of large extent or confined by surrounding cells, the {\em expected} concentration of \acp{NT} after a single release into the synaptic cleft is described by the following partial differential equation \cite{lotter21}
\begin{align}
    \partial_t c(x,t) = D\partial_{xx} c(x,t) - \ke c(x,t), \quad 0 < x < a,\label{eq:mean:rd}
\end{align}
where the synaptic cleft is represented by the one-dimensional spatial domain $[0,a]$, $c(x,t)$ denotes the expected concentration of solute \acp{NT} at time $t$ and spatial coordinate $x$ in $\si{\per\micro\meter}$, $D$ and $\ke$ denote the diffusion coefficient of the \acp{NT} in $\si{\micro\meter\squared\per\micro\second}$ and the enzymatic degradation rate in $\si{\per\micro\second}$, respectively, and $\partial_t$ and $\partial_{xx}$ denote the first partial derivative with respect to $t$ and the second partial derivative with respect to $x$, respectively.
The reversible binding of \acp{NT} to postsynaptic receptors is modeled as a boundary condition at $x=a$ \cite{lotter21}
\begin{equation}
    - D\,\partial_x c(x,t)\big\vert_{x=a} = \ka \left( 1 - \frac{o(t)}{C} \right) c(a,t) - \kd o(t),\label{eq:mean:sat_bdr}
\end{equation}
where $C$ and $o(t)$ denote the total number of postsynaptic receptors and the {\em expected} number of postsynaptic receptors occupied at time $t$, respectively, $\ka$ and $\kd$ denote the microscopic binding rate of \acp{NT} to postsynaptic receptors in $\si{\micro\meter\per\micro\second}$ and the unbinding rate of \acp{NT} from postsynaptic receptors in $\si{\per\micro\second}$, respectively, and $\partial_x$ denotes the first partial derivative with respect to $x$.
The model is completed by the initial and boundary conditions \cite{lotter21}
\begin{align}
    c(x,0) = N_0\delta(x) \quad\textrm{and}\quad \partial_x c(x,t)\big\vert_{x=0} = 0,\label{eq:mean:init_and_no_flux_bdr}
\end{align}
respectively, where $N_0$ and $\delta(\cdot)$ denote the number of released \acp{NT} and the Dirac delta distribution, respectively.
Furthermore, $o(t)$ is related to $c(x,t)$ by the equation
\begin{equation}
    o(t) = \int_{0}^{t} - D\,\partial_x c(x,\tau)\big\vert_{x=a} \dtau.\label{eq:mean:def_i}
\end{equation}

Since boundary condition \eqref{eq:mean:sat_bdr} is nonlinear, a closed-form solution to the boundary value problem \labelcref{eq:mean:rd,eq:mean:sat_bdr,eq:mean:init_and_no_flux_bdr} cannot be obtained.
Instead, a state space model is used in \cite{lotter21} to compute $o(t)$ iteratively in the spatio-temporal transform domain.
We call this model $\mathcal{S}$ and it is defined by a {\em state equation} \cite[Eq.~(42)]{lotter21} and an {\em output equation} \cite[Eq.~(31)]{lotter21}.

Now, let $n(t)$ denote the expected total number of \acp{NT}, i.e., the expected number of solute \acp{NT} and bound \acp{NT}, at time $t$.
When $\mathcal{S}$ is computed, we obtain not only $o(t)$, but also $c(x,t)$ and $n(t)$ \cite[Sec.~III-C-3]{lotter21}.
In Section~\ref{sec:system_model:macr_bd_rt}, we will use these quantities to compute the macroscopic absorption rate for \acp{NT} to postsynaptic receptors.

\subsection{The Postsynaptic Potential}\label{sec:system_model:psp}

As detailed in Section~\ref{sec:system_model:biological_background}, the activation of (ionotropic) postsynaptic receptors makes the membrane of the postsynaptic cell permeable for positively charged ions.
For most ionotropic receptors, such as \acp{AMPAR} and \acp{nAChR}, the relationship between ionic current flowing through the receptor and the postsynaptic membrane potential is linear\footnote{A notable exception, i.e., an ionotropic receptor with nonlinear current-voltage relationship is the \ac{NMDAR}.} \cite{byrne14}.
Hence, for such receptors, the ionic current flowing through a receptor in the activated state at time $t$, $\Isc(t)$ in \si{\pico\ampere}, can be written according to Ohm's law as \cite{byrne14}
\begin{align}
    \Isc(t) = \gsc (\Vm(t) - \Er),
\end{align}
where $\gsc$ denotes the single-channel conductance of the receptor in \si{\nano\siemens}, $\Vm(t)$ denotes the \ac{PSP} at time $t$ in \si{\milli\volt}, and $\Er$ denotes the {\em reversal potential} corresponding to the ion species for which the receptor is permeable in \si{\milli\volt} \cite{mccormick14}.
Consequently, the total synaptic current due to the random activation of ionotropic receptors, $\Isyn(t)$, is given as follows
\begin{align}
    \Isyn(t) = \gsc O(t) (\Vm(t) - \Er),\label{eq:def:Isyn}
\end{align}
where $O(t)$ denotes the random number of activated postsynaptic receptors at time $t$.

On the other hand, the postsynaptic membrane is permeable to some ions and ion pumps transport ions from the intracellular space to the extracellular space, cf.~Section~\ref{sec:system_model:biological_background}.
The ionic current caused by these properties of the membrane is called {\em leakage current}, denoted by $\IL(t)$ in \si{\pico\ampere}, and can be written in terms of the leakage conductance of the membrane $\gL$ in \si{\nano\siemens\per\square\micro\meter} and the leakage potential $\EL$ as \cite{byrne14}
\begin{align}
    \IL(t) = \gL (\Vm(t) - \EL).\label{eq:def:IL}
\end{align}

\begin{figure}[!tbp]
    \centering
    \includegraphics[width=\linewidth]{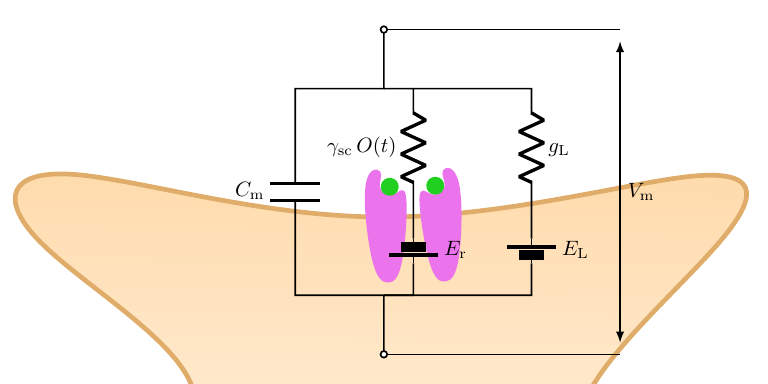}
    \caption{Equivalent circuit model for the \ac{PSP}. Ohmic currents across the membrane are due to diffusion of ions along their electrochemical gradients. The membrane separates charges and, hence, acts as capacitor.}
    \label{fig:psp_circuit}
\end{figure}

From \eqref{eq:def:Isyn} and \eqref{eq:def:IL} and the fact that the postsynaptic membrane constitutes a diffusion barrier for ions follows the equivalent circuit model for the postsynaptic membrane depicted in Fig.~\ref{fig:psp_circuit}.
According to this model, the \ac{PSP}, $\Vm(t)$, is determined by the following ordinary differential equation \cite{byrne14,destexhe01}
\begin{align}
    \Cm \frac{\textrm{d} \Vm(t)}{\textrm{d} t} &= - \IL(t) - \frac{1}{\A}\Isyn(t)\nonumber\\
    &= - \gL (\Vm(t) - \EL) - \frac{\gsc}{\A} O(t) (\Vm(t) - \Er),\label{eq:psp}
\end{align}
where $\Cm$ denotes the capacitance of the postsynaptic membrane in \si{\femto\farad} and $\A$ denotes the membrane surface area in \si{\micro\meter\squared}.
Considering $\Vm(t)$ as the output signal and $O(t)$ as the input signal, \eqref{eq:psp} constitutes a nonlinear filter for the non-stationary random process $O(t)$.
In the remainder of this section, a statistical model for the input process $O(t)$ in terms of the \ac{CME} is derived.

\subsection{Macroscopic Binding Rate}\label{sec:system_model:macr_bd_rt}

In the deterministic model \labelcref{eq:mean:rd,eq:mean:sat_bdr,eq:mean:init_and_no_flux_bdr,eq:mean:def_i}, the binding rate of the \acp{NT} to postsynaptic receptors is given by constant $\ka$.
In fact, $\ka$ results from a technique termed {\em boundary homogenization} \cite{berezhkovskii04} applied when mapping the actual three-dimensional reaction-diffusion process to the one-dimensional\footnote{By ``one-dimensional'', we refer to spatial dimensions, excluding the temporal dimension.} process in \labelcref{eq:mean:rd,eq:mean:sat_bdr,eq:mean:init_and_no_flux_bdr,eq:mean:def_i} \cite{lotter20}.
According to \cite[Sec.~V-A]{lotter21}, $\ka$ can be written as
\begin{equation}
    \ka = \kadt C,
\end{equation}
where $\kadt$ is a constant depending on the intrinsic binding rate of one \ac{NT} to one receptor and the ratio of the receptor area to the postsynaptic membrane surface area.
Hence, the activation of postsynaptic receptors can be written in terms of the following reversible bi-molecular reaction \eqref{eq:mean:sat_bdr}
\begin{equation}
    S_a + R \xrightleftharpoons[\kd]{\kadt} O,\label{eq:reac:intr_bd_rt}
\end{equation}
where $R$ denotes unoccupied postsynaptic receptors, $S_a$ denotes the solute \acp{NT} located in an (infinitesimally) small volume close to the postsynaptic membrane, and $O$ denotes activated postsynaptic receptors as defined in Section~\ref{sec:system_model:psp}.

Now, denoting by $S(t)$ the total number of solute molecules at time $t$ and assuming that the ratio $S_a(t)/S(t)$ is well-approximated by the ratio of the corresponding mean values obtained from \labelcref{eq:mean:rd,eq:mean:sat_bdr,eq:mean:init_and_no_flux_bdr,eq:mean:def_i}, i.e.,
\begin{align}
    \frac{S_a(t)}{S(t)} \approx \frac{\mathbb{E}[S_a(t)]}{\mathbb{E}[S(t)]} = \frac{c(a,t)}{\int_{0}^{a}c(x,t)\mathrm{d}x},\label{eq:diff_approx}
\end{align}
where $\mathbb{E}[\cdot]$ denotes the expectation operator, we can express the change in $O$ due to the binding and unbinding of \acp{NT} in the large system limit as follows
\begin{align}
    \frac{\mathrm{d}O(t)}{\mathrm{d}t} &= \kadt S_a(t) R(t) - \kd O(t)\nonumber\\ &\approx \kadt S(t) \frac{c(a,t)}{\int_{0}^{a}c(x,t)\mathrm{d}x} R(t) - \kd O(t).
\end{align}
Hence, defining the time-dependent macroscopic binding rate $\ka(t)$ as $\ka(t) = \kadt c(a,t)/\int_{0}^{a}c(x,t)\mathrm{d}x$, we obtain the reaction
\begin{equation}
    S + R \xrightleftharpoons[\kd]{\ka(t)} O.\label{eq:reac:macr_bd_rt}
\end{equation}

Eq.~\eqref{eq:reac:macr_bd_rt} provides a space-independent description of the reaction of \acp{NT} with postsynaptic receptors.
However, in contrast to space-independent models with constant reaction rates \cite{munsky18}, we do {\em not} assume that the reaction volume is well-mixed.
Instead, the spatially heterogeneous and time-dependent distribution of solute \acp{NT} is represented by $\ka(t)$.
The accuracy of this model as compared to the actual reaction-diffusion process depends on the validity of \eqref{eq:diff_approx}.
Eq.~\eqref{eq:diff_approx} in turn is justified if the number of solute \acp{NT} is large compared to the size of the synapse and diffusion is relatively fast as compared to the chemical reactions.
We will show in Section~\ref{sec:numerical_results} that \eqref{eq:reac:macr_bd_rt} provides a very accurate model for different, biologically relevant ranges of parameter values.

\subsection{The Chemical Master Equation}\label{sec:system_model:cme}

In this section, we formulate a statistical model for the random processes governing the activation of postsynaptic receptors and the degradation of solute \acp{NT} in terms of the \ac{CME}.
To this end, we denote the random total number of \acp{NT}, i.e., the number of solute \acp{NT} and bound \acp{NT}, at time $t$ as $N(t)$ and recall from Section~\ref{sec:system_model:macr_bd_rt} that the random number of activated receptors at time $t$ is denoted by $O(t)$.

First, besides the reaction of \acp{NT} with postsynaptic receptors defined in \eqref{eq:reac:macr_bd_rt}, solute \acp{NT} are exposed to enzymatic degradation which is modeled as a uni-molecular reaction in \eqref{eq:mean:rd}.
This degradation reaction is described as follows
\begin{align}
    S \xrightharpoonup{\ke} \varnothing,\label{eq:reac:enz_deg}
\end{align}
where $\varnothing$ denotes any species that does not react with \acp{NT} and postsynaptic receptors.

Next, we note that the state of the system described by \labelcref{eq:reac:macr_bd_rt,eq:reac:enz_deg} at time $t$ is fully determined by random variables $N(t)$ and $O(t)$, since $S(t) = N(t) - O(t)$ and $R(t) = C - O(t)$.
Furthermore, denoting the time-dependent joint probability mass function of $N(t)$ and $O(t)$ as $\P(n,o,t)$, the time-evolution of $\P(n,o,t)$ is governed by the \ac{CME} \eqref{eq:cme} on the top of the next page, where $n \in \left\lbrace 0,\ldots,N_0 \right\rbrace$, $o \in \left\lbrace 0,\ldots,C \right\rbrace$.
\begin{figure*}
    \begin{align}
        \frac{\partial \P(n, o, t)}{\partial t} = &-\left[\kd o + \ke (n-o) + \ka(t + t_0)(n-o)(C-o)\right]\P(n,o,t) + \kd(o+1)\P(n,o+1,t) \nonumber\\
        &{} + \ke (n + 1 - o) \P(n+1,o,t) + \ka(t+t_0) (n-o+1)(C-o+1)\P(n,o-1,t)\label{eq:cme}
    \end{align}        
    \hrulefill
\end{figure*}
By specifying a deterministic initial value $(n_0,o_0)$ for \eqref{eq:cme}, i.e.,
\begin{align}
    \P(n, o, t_0) = \begin{cases}
        1,\qquad&\textrm{if}\,(n,o) = (n_0,o_0),\\
        0,\qquad&\textrm{otherwise},
    \end{cases}
\end{align}
we obtain 
\begin{align}
    \P(n, o, t) &= \Pr\lbrace (N(t_0+t),O(t_0+t)) = (n,o) | \nonumber\\
    &\qquad  (N(t_0),O(t_0)) = (n_0,o_0) \rbrace.
\end{align}
Finally, to define \eqref{eq:cme} on the boundary of the state space, we set $\P(-1, \cdot, \cdot) \equiv 0$, $\P(N_0+1, \cdot, \cdot) \equiv 0$, $\P(\cdot, -1, \cdot) \equiv 0$, and $\P(\cdot, C+1, \cdot) \equiv 0$.

\begin{figure}[t]
    \centering
    \includegraphics[width=.75\linewidth]{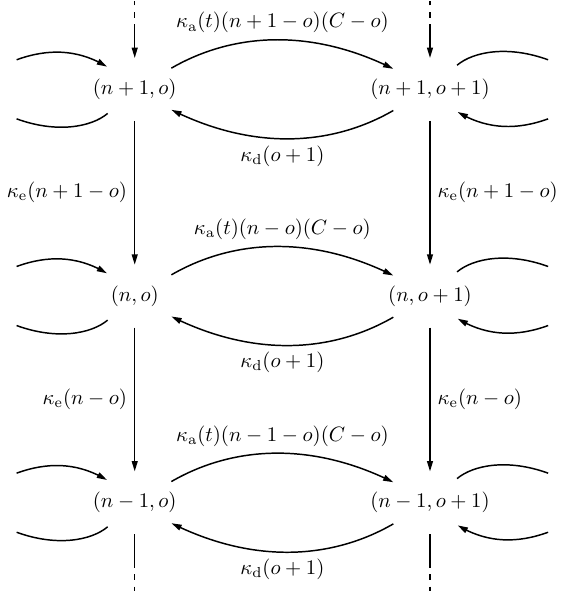}
    \caption{State transitions modeled by the \ac{CME} \eqref{eq:cme}. Horizontal state transitions correspond to the reversible reaction of \acp{NT} with postsynaptic receptors. Vertical state transitions correspond to the enzymatic degradation of solute \acp{NT}.}
    \label{fig:state_transitions}
\end{figure}

Equation~\eqref{eq:cme} defines a discrete-state random process.
Since the state transition probabilities are time-dependent, this process is not strictly Markovian, as the {\em waiting times} in each state are not exponentially distributed \cite{castro18}.
However, since the state transitions in \eqref{eq:cme} depend only on the current state of the process and the absolute time, the process $(N(t),O(t))$ still fulfills a Markov property of the form
\begin{multline}
    \Pr\lbrace\left. (N(t_2),O(t_2)) \,\right| (N(t_1),O(t_1)), (N(t_0),O(t_0)) \rbrace\\ = \Pr\lbrace\left. (N(t_2),O(t_2)) \,\right| (N(t_1),O(t_1)) \rbrace,\label{eq:markov}
\end{multline}
for any $t_2 > t_1 > t_0$.
The state transitions corresponding to \eqref{eq:cme} are illustrated in Fig.~\ref{fig:state_transitions}.

\begin{figure}[t]
    \includegraphics[width=\linewidth]{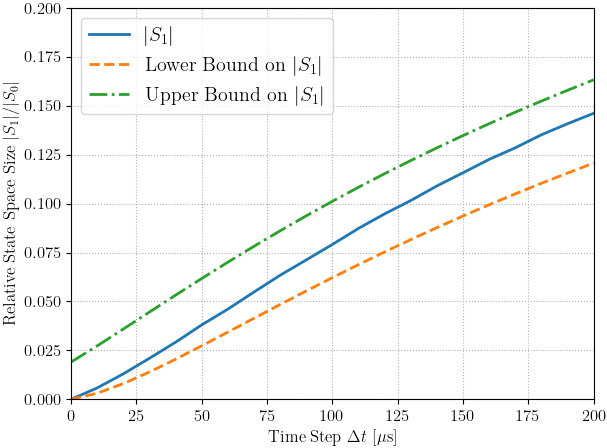}
    \caption{Size of the reduced state space $S_1$ relative to the size of the full state space $S_0$ for different values of $\Delta t$ (solid line). The lower and upper bounds on $|S_1|$ resulting from \eqref{eq:bound_N_min} and \eqref{eq:bound_O_max} are shown as dashed and dash-dotted lines, respectively.}
    \label{fig:ss_analysis_dt}
\end{figure}

\section{Postsynaptic Membrane Potential}
\label{sec:psp}
In this section, we first present an approximation for \eqref{eq:psp} in terms of a linear filter.
Based on this approximation, the \ac{PSP} is given as a function of the activation of postsynaptic receptors represented by random process $O(t)$.
In the second part of this section, we derive expressions for the mean and the variance of the \ac{PSP} in terms of the mean and the autocovariance of $O(t)$.
Finally, we propose an approximation of the instantaneous statistics of the \ac{PSP} based on the Gaussian distribution.

\subsection{Linear Approximation of the PSP}\label{sec:psp:linear_approximation}

We recall that the \ac{PSP} $\Vm(t)$ is defined by the linear, first-order differential equation \eqref{eq:psp}.
Since \eqref{eq:psp} represents a nonlinear filter of the input signal $O(t)$ and since this nonlinearity complicates the evaluation of the statistics of $\Vm(t)$, we seek an approximation of \eqref{eq:psp} in terms of a linear filter.
Fortunately, such an approximation is indeed possible and justified in biologically plausible parameter ranges.

The postsynaptic membrane acts as a nonlinear filter, because the activation of postsynaptic receptors leads to a change of the {\em conductance} of the postsynaptic membrane.
In other words, the current due to the {\em input signal} $O(t)$, $\Isyn(t)$, depends on the {\em output signal} $\Vm(t)$, cf.~\eqref{eq:def:Isyn}, and this renders the filter \eqref{eq:psp} nonlinear.
Now, the postsynaptic membrane is polarized to the leakage potential $\EL \approx \SI{-80}{\milli\volt}$ under resting conditions and gets depolarized towards the reversal potential $\Er \approx \SI{0}{\milli\volt}$ when postsynaptic receptors are activated \cite{byrne14}.
However, the magnitude of the \ac{PSP} after synaptic transmission in one single synapse is only around $\SI{1}{\milli\volt}$, i.e., $|\EL-\Vm(t)| \approx \SI{1}{\milli\volt}$, which is small compared to $|\Vm(t)-\Er|$ \cite{byrne14}.
Hence, with respect to $\Isyn(t)$, $\Vm(t)$ can be approximated as $\Vm(t) \approx \EL$.
This leads to the following linear approximation of \eqref{eq:psp}
\begin{align}
    \Cm \frac{\textrm{d} \Vmt(t)}{\textrm{d} t} = - \gL (\Vmt(t) - \EL) - \frac{\gsc}{\A} O(t) (\EL - \Er),\label{eq:psp_linear}
\end{align}
where $\Vmt(t)$ denotes the linear approximation of $\Vm(t)$.
Eq.~\eqref{eq:psp_linear} is an approximation for \eqref{eq:psp} which is commonly used in the literature \cite{brunel01}.
The accuracy of \eqref{eq:psp_linear} is further validated by the results presented in Section~\ref{sec:numerical_results}.

In order to solve \eqref{eq:psp_linear}, we change variables as $V(t) = \Vmt(t) - \EL$ and introduce the auxiliary variables $\alpha=\gL/\Cm$ and $\beta=\gsc(\Er-\EL)/(\A\Cm)$ to simplify the notation.
With these substitutions and assuming without loss of generality that the postsynaptic membrane is at rest at $t=0$, i.e., $V(0) = 0$, $V(t)$ is given as follows
\begin{align}
    V(t) = \beta \int_{0}^{t} \exp\left[-\alpha(t-\tau)\right] O(\tau) \dtau.\label{eq:psp_sol}
\end{align}

From \eqref{eq:psp_sol} and the definitions of $\alpha$ and $\beta$, it is evident how the physical parameters of the postsynaptic membrane shape the \ac{PSP}.
The ratio of the leakage conductance $\gL$ to the membrane capacitance $\Cm$, $\alpha$, determines how fast the \ac{PSP} decays after a synaptic event.
On the other hand, the single-channel conductance $\gsc$ and the difference between leakage potential and reversal potential $\Er-\EL$ relative to $\Cm$, i.e., $\beta$, determine by how much the activation of a given number of postsynaptic receptors depolarizes the postsynaptic membrane.

\subsection{Instantaneous Statistics of the PSP}

In the previous section, the random process $V(t)$ was defined as a function of the postsynaptic receptor activation $O(t)$.
Since $O(t)$ is a random process, however, the integral in \eqref{eq:psp_sol} is a stochastic integral \cite{schuss09} and can, hence, only be evaluated directly for specific realizations of $O(t)$.
To characterize $V(t)$ statistically, we first derive its mean and variance.

To this end, we define the {\em autocovariance} of $O(t)$ as a function of the time variables $t_1$ and $t_2$ as follows
\begin{align}
    \KOO(t_1,t_2) &= \Cov(O_{t_1},O_{t_2})\nonumber\\ &= \mathbb{E}\left[\left(O(t_1)-\mathbb{E}\left[O(t_1)\right]\right)\left(O(t_2)-\mathbb{E}\left[O(t_2)\right]\right)\right]\nonumber\\
    &= \mathbb{E}\left[O(t_1)O(t_2)\right] - \mathbb{E}\left[O(t_1)\right]\mathbb{E}\left[O(t_2)\right].\label{eq:def:cov}
\end{align}
Now, we state the main result of this section.

\begin{theorem}\label{thm:psp}
    The mean and the variance of $V(t)$ as defined in \eqref{eq:psp_sol} are given by
    \begin{align}
        v(t) &= \Exp{V(t)} = \beta \int_{0}^{t} \exp\left[-\alpha(t-\tau)\right] \Exp{O(\tau)} \dtau,\label{eq:psp:mean}\\
        \sigma^2_V(t) &= \textrm{Var}(V(t)) = \Exp{\left(V(t)-\Exp{V(t)}\right)^2}\nonumber\\
        &= \beta^2 \int_{0}^{t} \int_{0}^{t} \exp\left[-\alpha(t-\tau_1)\right] \exp\left[-\alpha(t-\tau_2)\right]\,\nonumber\\
        &\quad \times \KOO(\tau_1,\tau_2)\,\dtau_2\,\dtau_1,\label{eq:psp:var}
    \end{align}
    respectively, where $\KOO(t_1,t_2)$ was defined in \eqref{eq:def:cov}.
\end{theorem}
\begin{IEEEproof}
    See Appendix~\ref{sec:app:prf_thm_psp}.
\end{IEEEproof}

We note from \eqref{eq:psp_sol} (and even more explicitly from the proof of Theorem~\ref{thm:psp}) that the \ac{PSP} is ultimately an accumulation of random states which in turn results from the accumulation of many individual random events, namely the binding and unbinding of \acp{NT} to and from postsynaptic receptors, respectively.
Although these binding events are partially correlated, as we will see in Section~\ref{sec:numerical_results}, we argue that the statistical dependence is small enough compared to the time scale on which the filter \eqref{eq:psp} operates such that the central limit theorem applies here.
Hence, we propose to model the instantaneous statistics of $\Vm(t)$ as follows
\begin{equation}
    \Vm(t) \sim \mathcal{N}(v(t) + \EL, \sigma^2_V(t)),\label{eq:psp:gaussian_approx}
\end{equation}
where $\mathcal{N}(\mu, \sigma^2)$ denotes the Gaussian distribution with mean $\mu$ and variance $\sigma^2$.
The accuracy of this approximation is verified in Section~\ref{sec:numerical_results}.

\section{Solving the Chemical Master Equation}    
\label{sec:cme}
A closed-form solution of the system of equations specified by \eqref{eq:cme} is in general not possible \cite{schnoerr17}.
Hence, in this section, we first aim at computing $\P(n,o,t)$ as defined in \eqref{eq:cme} numerically.
As we will see, even the numerical evaluation of \eqref{eq:cme} poses a severe challenge.
We will then leverage the proposed method towards the end of this section to compute the autocovariance $\KOO(t_1,t_2)$ of the random process $O(t)$.
According to Theorem~\ref{thm:psp}, this will finally allow us to characterize the variance of the \ac{PSP}.

\subsection{Formal Solution}\label{sec:cme:formal_solution}

According to \eqref{eq:cme}, there exist $M=(N_0+1)\times(C+1)$ different system states.
We organize these states in a level-dependent manner where the total number of \acp{NT}, $n$, determines the level.
Accordingly, we define the probability vector $\pibd(t) \in [0,1]^{M}$ as follows
\begin{equation}
    \pibd(t) = [\pibd_{N_0}(t), \pibd_{N_0-1}(t), \ldots, \pibd_{0}(t)]^{\mathrm{T}},\label{eq:def:pibd}
\end{equation}
where $[\cdot]^{\mathrm{T}}$ denotes transposition and the $N_0+1$ vectors $\pibd_{n}(t) \in [0,1]^{(C+1)}$ are defined as\footnote{Note that we allow for infeasible states in this definition, since $P(n,o,\cdot) \equiv 0$ for $n < o$. This is done only for notational simplicity, infeasible states are omitted in all practical computations.} $\pibd_{n}(t) = [P(n,0,t),P(n,1,t),\ldots,P(n,C,t)]$.

In a similar fashion, we collect all transition probabilities from \eqref{eq:cme} in the time-dependent transition matrix $\Abd(t) \in \mathbb{R}^{M \times M}$.
$\Abd(t)$ is a block-bidiagonal matrix consisting of $(N_0+1)^2$ $(C+1) \times (C+1)$ matrices with all sub-matrices equal to the $(C+1) \times (C+1)$ all-zero matrix $\boldsymbol{0}_{(C+1)}$, except for the matrices on the main diagonal and the lower diagonal which we denote as
\begin{equation}
    \Abd_{i,i} = \Qbd_{N_0-i+1},\quad 1 \leq i \leq N_0+1,\label{eq:A:Q}
\end{equation}
and
\begin{equation}
    \Abd_{i,i-1} = \Dbd_{N_0-i+1},\quad 2 \leq i \leq N_0+1,\label{eq:A:D}
\end{equation}
respectively.
Matrices $\Qbd_{n}$ and $\Dbd_{n}$ collect the level-dependent transition rates for the binding and degradation reactions, respectively.
The $\Qbd_{n}$ are tridiagonal matrices with the diagonal elements defined as follows
\begin{align}
    &\left(\Qbd_{n}\right)_{i+1,i+1}&=\,& -\left[\kd i + \ke (n-i) \right.\nonumber\\
    &&&\left. + \ka(t+t_0)(n-i)(C-i)\right],\,\, 0 \leq i \leq C,\\
    &\left(\Qbd_{n}\right)_{i+1,i}&=\,& \ka(t+t_0) (n-i+1)\nonumber\\
    &&& \times (C-i+1),\qquad\qquad\, 1 \leq i \leq C,\\
    &\left(\Qbd_{n}\right)_{i+1,i+2}&=\,& \kd (i+1),\qquad\qquad\,\, 0 \leq i \leq C-1.
\end{align}
The $\Dbd_{n}$ are diagonal matrices with the main diagonal elements defined as follows
\begin{equation}
    \left(\Dbd_{n}\right)_{i+1,i+1} = \ke (n + 1 - i),\qquad\qquad 0 \leq i \leq C.
\end{equation}
With these definitions, we rewrite \eqref{eq:cme} in vector form as the following system of differential equations
\begin{equation}
    \frac{\mathrm{d}\pibd(t)}{\dt} = \Abd(t)\pibd(t),\label{eq:cme:vec}
\end{equation}
the formal solution of which is given as follows
\begin{equation}
    \pibd(t) = \exp\left(\int_0^t \Abd(\tau) \dtau\right)\pibd_0,\label{eq:cme:vec:sol}
\end{equation}
where $\exp(\boldsymbol{M})$ denotes the matrix exponential of square matrix $\boldsymbol{M}$.
In the special case $t_0=0$, $\pibd_0 = \pibd(0)$ is given by the $M$-dimensional vector $[1,0,\ldots,0]^{\mathrm{T}}$, cf.~\eqref{eq:mean:init_and_no_flux_bdr}.

\subsection{Computational Issues and Approximation Methods}\label{sec:cme:computational_issues}

Since the dimension of $\Abd(t)$ grows quadratically with both the number of released \acp{NT} and the number of receptors, computing the matrix exponential in \eqref{eq:cme:vec:sol} is intractable \cite{moler03}.
Indeed, even for a moderate number of $500$ released \acp{NT} and $200$ receptors, the number of elements of $\Abd(t)$ is of order $\sim 10^{10}$.

This problem is common to many applications using the \ac{CME} as modeling tool and, consequently, several methods have been proposed to approximate the solution of the \ac{CME} \cite{schnoerr17}.
Two of the most frequently used approximation methods are {\em moment closure} schemes and schemes exploiting some kind of {\em system size expansion}, the most popular among the latter being the {\em \ac{LNA}} \cite{schnoerr17}.
Both of these approaches have their strengths and limitations, the detailed discussion of which would go far beyond the scope of this paper.
Here, it suffices to say that due to the bimolecular reaction \eqref{eq:reac:macr_bd_rt} both methods cannot be used to obtain the statistics of $N(t)$ and $O(t)$ without further simplifications or approximations.

Another commonly used method for computing high-dimensional \acp{CME} is to approximate the \ac{CME} on a lower-dimensional subspace of its state space\footnote{Such state reduction schemes are also referred to as {\em state lumping schemes} \cite{munsky18}.}.
Classical state reduction schemes for the \ac{CME} operate on a reduced but static state space, meaning the state space does not change over time \cite{munsky18}.
In the following section, we show how to exploit our knowledge of the first-order statistics of $N(t)$ and $O(t)$ given by $n(t)$ and $o(t)$, respectively, to adapt the state space iteratively while computing the \ac{CME}.
We show that this adaptive scheme allows to compute \eqref{eq:cme:vec:sol} efficiently and, at the same time, control the approximation error.

\subsection{Adaptive State Reduction}\label{sec:cme:state_reduction}

To introduce the proposed adaptive state reduction scheme, we first discretize time into subsequent intervals of length $\Delta t$, such that the $k$th interval is $I_k = [t_k,t_{k+1}]$, where $t_k = (k-1) \Delta t$ and $k$ is from the set of positive integers $\mathbb{N}$.
The idea is to compute $\pibd(t)$ iteratively for each interval $k$ while discarding the states $(n,o)$ which do not contribute significant probability mass in interval $k$.

To this end, we first define the respective marginal distributions of $N(t)$ and $O(t)$ at time $t$ as follows
\begin{equation}
    P_N(n,t) = \sum_{o=0}^{C} P(n,o,t) \,\,\textrm{and}\,\, P_O(o,t) = \sum_{n=0}^{N_0} P(n,o,t),
\end{equation}
and the {\em full state space} of \eqref{eq:cme} as
\begin{equation}
    S_0 = \lbrace (n,o)| 0 \leq n \leq N_0, 0 \leq o \leq C \rbrace.\label{eq:def:S_0}
\end{equation}
Furthermore, let $P_B(\cdot;n,p)$ denote the probability mass function of a binomial random variable with parameters $n$ and $p$, and define
\begin{align}
    \Nmin^{(k)} &= \max \left\lbrace n \left| \sum_{n'=0}^{n} P_B\left(n';N_0,\frac{n(t_{k+1})}{N_0}\right)\right. < \epsilon \right\rbrace,\label{eq:def:Nmin}\\
    \Nmax^{(k)} &= \min \left\lbrace n\, \left| \sum_{n'=n}^{N} P_N\left(n',t_k\right) < \epsilon \right\rbrace\right.,\label{eq:def:Nmax}\\
    \Omin^{(k)} &= \max \left\lbrace o\, \left| \max_{t \in I_k} \sum_{o'=0}^{o} P_B\left(o';C,\frac{o(t)}{C}\right) < \epsilon \right\rbrace\right.,\label{eq:def:Omin}\\
    \Omax^{(k)} &= \min \left\lbrace o\, \left| \max_{t \in I_k} \sum_{o'=o}^{C} P_B\left(o';C,\frac{o(t)}{C}\right) < \epsilon \right\rbrace\right.,\label{eq:def:Omax}
\end{align}
where $\epsilon > 0$ denotes a threshold parameter for discarding states that with high probability do not occur in interval $k$.
Hence, \labelcref{eq:def:Nmin,eq:def:Nmax,eq:def:Omin,eq:def:Omax} provide estimates for the minimum and maximum numbers of \acp{NT} and activated receptors, respectively, that may be observed in interval $k$.

We define the {\em reduced state space} in interval $k$ as follows
\begin{equation}
    S_k = \lbrace (n,o)| \Nmin^{(k)} \leq n \leq \Nmax^{(k)}, \Omin^{(k)} \leq o \leq \Omax^{(k)} \rbrace,\label{eq:def:S_k}
\end{equation}
and the restriction of $\pibd(t)$ to $S_k$ as
\begin{equation}
    \pibd(t)|_{S_k} = [\pibd_{(\Nmax^{(k)})}(t)|_{S_k}, \ldots, \pibd_{(\Nmin^{(k)})}(t)|_{S_k}]^{\mathrm{T}},\label{eq:def:pibd_S_k}
\end{equation}
where
\begin{equation}
    \pibd_{n}(t)|_{S_k} = [P(n,\Omin^{(k)},t),\ldots,P(n,\Omax^{(k)},t)].
\end{equation}
The restriction of $\Abd(t)$ to $S_k$, $\Abd(t)|_{S_k}$, is obtained by discarding the rows and the columns of $\Abd(t)$ corresponding to the indices of the elements of $\pibd(t)$ discarded in $\pibd(t)|_{S_k}$.
Finally, we define the approximate solution of \eqref{eq:cme:vec} in interval $I_k$, $\hat{\pibd}^{(k)}(t)$, as the solution of the following system of equations
\vspace*{-1mm}
\begin{equation}
    \frac{\mathrm{d} \hat{\pibd}^{(k)}(t)}{\mathrm{d} t} = \Abd(t)|_{S_k} \hat{\pibd}^{(k)}(t),\label{eq:def:pibd_hat}
\end{equation} 
where $t \in I_k$ and $\hat{\pibd}^{(k)}(t_k) = \pibd(t_k)|_{S_k}$.

The following theorem justifies these definitions.
\begin{theorem}\label{thm:state_reduction}
    Let $k \in \mathbb{N}$ and assume $\pibd(t_k)$ is known. Then, for any $\epsilon > 0$,
    \begin{equation}
        ||\pibd(t)|_{S_k} - \hat{\pibd}^{(k)}(t)||_1 < 4 \epsilon, \quad\forall\, t \in I_k,
    \end{equation}
    where $\pibd(t)|_{S_k}$, $S_k$, and $\hat{\pibd}^{(k)}(t)$ are defined in \eqref{eq:def:pibd_S_k}, \eqref{eq:def:S_k}, and \eqref{eq:def:pibd_hat}, respectively, and $||\boldsymbol{v}||_1$ denotes the $l_1$ norm of vector $\boldsymbol{v}$.\footnote{State space $S_k$ depends on the choice of $\epsilon$ by definitions \labelcref{eq:def:Nmin,eq:def:Nmax,eq:def:Omin,eq:def:Omax,eq:def:S_k}. This dependence remains implicit for notational simplicity.}
\end{theorem}
\begin{IEEEproof}
    See Appendix~\ref{sec:app:prf_thm_state_reduction}.
\end{IEEEproof}

Theorem~\ref{thm:state_reduction} allows us to approximate the solution to the \ac{CME} \eqref{eq:cme} by iteratively solving the lower-dimensional problem \eqref{eq:def:pibd_hat} for each interval $k$.
To state the iterative algorithm, we need yet to define how to map $\hat{\pibd}^{(k)}(t)$ to $\hat{\pibd}^{(l)}(t)$ for any $k,l \in \mathbb{N} \bigcup \lbrace 0 \rbrace$.
To this end, let us denote the elements of $\hat{\pibd}^{(k)}(t)$ by
\begin{equation}
    \hat{\pibd}^{(k)}(t) = \left[\hat{P}^{(k)}(n_{i_1},o_{i_1},t),\ldots,\hat{P}^{(k)}(n_{i_{|S_{k}|}},o_{i_{|S_{k}|}},t)\right]^{\mathrm{T}},
\end{equation}
where the indices $i_1,\ldots,i_{|S_{k}|}$ enumerate the states in state space $S_{k}$ and $|S_{k}|$ denotes the number of states in $S_{k}$.
We define now the projection of $\hat{\pibd}^{(k)}(t)$ onto state space $S_{l}$ as follows
\begin{equation}
    \mathcal{P}_{k \to l} \hat{\pibd}^{(k)}(t) = \left[\bar{P}^{(l)}(n_{j_1},o_{j_1},t),\ldots,\bar{P}^{(l)}(n_{j_{|S_{l}|}},o_{j_{|S_{l}|}},t)\right]^{\mathrm{T}},
\end{equation}
where the $j_1,\ldots,j_{|S_{l}|}$ enumerate the states in state space $S_{l}$, and
\begin{equation}
    \bar{P}^{(l)}(n_{j_m},o_{j_m},t) = \begin{cases}
        \hat{P}^{(k)}(n_{j_m},o_{j_m},t),&(n_{j_m},o_{j_m}) \in S_k,\\
        0,&\textrm{otherwise}.
    \end{cases}
\end{equation}

The proposed adaptive state reduction algorithm solves \eqref{eq:def:pibd_hat} and then maps the result to the reduced state space of the next interval in an iterative manner.
The complete algorithm is presented as Algorithm~\ref{alg:cme_lumped_state_space} at the top of this page.

\begin{algorithm}[t]
    \small
    \caption{Iterative computation of $\pibd(t)$}
    \begin{algorithmic}[1]\label{alg:cme_lumped_state_space}
        \STATE \textbf{input:} $t_0$, $\pibd_0$, $\Delta t$, $\epsilon$.
        \STATE \textbf{initialize:} $k=1$, $K=\left\lceil t/\Delta t \right\rceil$, $\hat{\pibd}^{(0)}(0) = \pibd_0$.
        \WHILE{$k \leq K$}
        \STATE Compute $S_k$ according to \labelcref{eq:def:Nmin,eq:def:Nmax,eq:def:Omin,eq:def:Omax,eq:def:S_k}.
        \STATE Set $\hat{\pibd}^{(k)}(t_{k}) = \mathcal{P}_{k-1 \to k} \hat{\pibd}^{(k-1)}(t_{k})$.
        \STATE Compute $\hat{\pibd}^{(k)}(t)$ for $t \in I_k$ by solving \eqref{eq:def:pibd_hat}.
        \STATE Set $k=k+1$.
        \ENDWHILE
        \RETURN $\mathcal{P}_{K \to 0}\hat{\pibd}^{(K)}(t)$.
    \end{algorithmic}
\end{algorithm}

\vspace{-4mm}
\subsection{Computational Efficiency of Algorithm 1}

In this section, we confirm the computational efficiency of Algorithm~\ref{alg:cme_lumped_state_space} as compared to solving \eqref{eq:cme:vec} directly, i.e., computing \eqref{eq:cme:vec:sol}.

The computational costs for solving \eqref{eq:cme:vec} and Algorithm~\ref{alg:cme_lumped_state_space} are dominated by the matrix exponentials $\exp\left(\int_0^t \Abd(\tau) \dtau\right)$ and $\exp\left(\int_{t_k}^{t_{k+1}} \Abd(\tau)|_{S_k} \dtau\right)$, respectively.
Since both matrices $\Abd(t)$ and $\Abd(t)|_{S_k}$ are sparse, the computational complexity of computing these matrix exponentials is proportional to $|S_0|^2$ and $|S_k|^2$, respectively \cite{almohy11}.
Hence, in order to compare the costs of computing \eqref{eq:cme:vec:sol} and Algorithm~\ref{alg:cme_lumped_state_space}, respectively, it is sufficient to compare $|S_0|$ and $|S_k|$.

According to \labelcref{eq:def:Nmin,eq:def:Nmax,eq:def:Omin,eq:def:Omax,eq:def:S_k}, the $|S_k|$ depend on the choice of $\Delta t$ and $\epsilon$, as well as on $n(t)$ and $o(t)$.
Now, to facilitate the presentation, we assume $k=1$, $t_0 = 0$, and $\Delta t < \argmax_t o(t)$, which implies $\Nmax^{(k)} = \Nmax^{(1)} = N_0$ and $\Omin^{(k)} = \Omin^{(1)} = 0$.\footnote{This assumption is not restrictive, since the arguments used in the following to bound $|S_1|$ can as well be developed to bound $|S_k|$ in the general case $k \in \mathbb{N}$.}

With the assumption just made, \eqref{eq:def:Omax} simplifies to $\Omax^{(1)} = \min \left\lbrace o\, \left| \sum_{o'=o}^{C} P_B\left(o';C,\frac{o(\Delta t)}{C}\right) < \epsilon \right\rbrace\right.$ and $|S_1| = \left(\Omax^{(1)}+1\right)\left(N_0-\Nmin^{(1)}+1\right)$.
To estimate $\Nmin^{(1)}$ and $\Omax^{(1)}$, we apply the following tail bounds for the binomial distribution \cite[Ch.~4, Eq.~(4.7.2)]{ash12}
\begin{multline}
    \frac{1}{\sqrt{2 n}}\exp\left(-n \DKL\left(\frac{m}{n}||p\right)\right) \\
    \leq \sum_{m'=0}^{m} P_B(m';n,p) \leq \exp\left(-n \DKL\left(\frac{m}{n}||p\right)\right),\label{eq:binom_tail_bound}
\end{multline}
where
$\DKL(p||q)$ denotes the Kullback-Leibler divergence in nats, i.e., $\DKL(p||q) = p \ln\left(\frac{p}{q}\right) + (1-p) \ln\left(\frac{1-p}{1-q}\right)$.
Furthermore, using the following bounds on $\DKL(p||q)$ \cite{vanerven14}
\begin{equation}
    2 |p-q|^2 \leq \DKL(p||q) \leq \ln\left(\frac{p^2}{q} + \frac{(1-p)^2}{(1-q)}\right),\label{eq:dkl_bounds}
\end{equation}
we finally obtain from \eqref{eq:def:Nmin} and \eqref{eq:def:Omax}
\begin{multline}
    n(\Delta t) - \sqrt{\frac{N_0}{2}\ln\left(\epsilon^{-1}\right)} \leq \Nmin^{(1)} \leq n(\Delta t)\\
     - \sqrt{\left(N_0-n(\Delta t)\right)n(\Delta t)\left(\left(\epsilon\sqrt{2N_0}\right)^{-1/N_0}-1\right)},\label{eq:bound_N_min}
\end{multline}
and 
\begin{multline}
    o(\Delta t) + \sqrt{\left(C-o(\Delta t)\right)o(\Delta t)\left(\left(\epsilon\sqrt{2C}\right)^{-1/C}-1\right)}\\
    \leq \Omax^{(1)} \leq o(\Delta t) + \sqrt{\frac{C}{2}\ln\left(\epsilon^{-1}\right)},\label{eq:bound_O_max}
\end{multline}
respectively.

Inequalities \eqref{eq:bound_N_min} and \eqref{eq:bound_O_max} show that the mean values $n(\Delta t)$ and $o(\Delta t)$ dominate $\Nmin^{(1)}$ and $\Omax^{(1)}$ and, therefore, $|S_1|$.
In contrast, in terms of the threshold parameter $\epsilon$, $|S_1|$ grows at most logarithmically in $\epsilon^{-1}$.
Hence, \eqref{eq:bound_N_min} and \eqref{eq:bound_O_max} indicate that $\Delta t$ should be chosen carefully, while the computational complexity of Algorithm~\ref{alg:cme_lumped_state_space} is less sensitive towards the choice of $\epsilon$.

To further elucidate how $|S_1|$ and $|S_0|$ relate to each other quantitatively, Fig.~\ref{fig:ss_analysis_dt} shows $|S_1|/|S_0|$ for different values of $\Delta t$ as defined by \labelcref{eq:def:Nmin,eq:def:Nmax,eq:def:Omin,eq:def:Omax,eq:def:S_k} and predicted by \eqref{eq:bound_N_min} and \eqref{eq:bound_O_max}, respectively.
The results presented in Fig.~\ref{fig:ss_analysis_dt} were obtained for the default parameter values given in Tables~\ref{tab:parameter_values} and \ref{tab:parameter_values_psp}.
First, we observe from Fig.~\ref{fig:ss_analysis_dt} that inequalities \eqref{eq:bound_N_min} and \eqref{eq:bound_O_max} provide indeed a useful characterization of $|S_1|$.
Furthermore, Fig.~\ref{fig:ss_analysis_dt} indicates that the proposed state reduction scheme leads to a reduction of the state space size by more than $95\%$ compared to $|S_0|$ for $\Delta t = \SI{50}{\micro \second}$ (the default value of $\Delta t$ used in this paper).
Since the computational complexity of Algorithm~\ref{alg:cme_lumped_state_space} scales with the square of the state space size, it is therefore reduced by more than $99.75\%$ as compared to the complexity of the original \ac{CME} problem.
At the same time, \eqref{eq:bound_N_min} and \eqref{eq:bound_O_max} assert that the computational cost of Algorithm~\ref{alg:cme_lumped_state_space} is rather insensitive towards the threshold parameter $\epsilon$ and, hence, high accuracy can be achieved without compromising computational efficiency.
This confirms the efficiency of the proposed state reduction scheme.

\subsection{Receptor Occupancy Autocovariance}\label{sec:cme:cov}

So far, we have discussed the statistical characterization of $O(t)$ and $N(t)$ in terms of their (joint) instantaneous distribution.
This means, we have computed $\P(n,o,t)$ for any time instant $t$.
In this section, we generalize the method developed in Section~\ref{sec:cme:state_reduction} to compute the autocovariance of $O(t)$, $K_{OO}(t_1,t_2)$, as defined in \eqref{eq:def:cov}.

First, we note that in order to compute $K_{OO}(t_1,t_2)$, the joint distribution of the random variables $O_{t_1}$ and $O_{t_2}$ which we denote by $P_O(o_1,o_2,t_1,t_2)$ is required.
Given any deterministic initial value $(N(t_0),O(t_0))=(n_0,o_0)$ and due to \eqref{eq:markov}, $P_O(o_1,o_2,t_1,t_2)$ is given by \eqref{eq:cov_joint_o} on the top of the next page.
\begin{figure*}
\begin{align}
    P_O(o_1,o_2,t_1,t_2) =& \sum_{n_1,n_2=0}^{N_0} \Pr\lbrace\left. (N(t_2),O(t_2)) = (n_2,o_2) \,\right| (N(t_1),O(t_1)) = (n_1,o_1) \rbrace\nonumber\\
    &{}\times \Pr\lbrace \left. (N(t_1),O(t_1)) = (n_1,o_1)\,\right| (N(t_0),O(t_0)) = (n_0,o_0) \rbrace.\label{eq:cov_joint_o}
\end{align}
\hrulefill    
\end{figure*}

Since $\Cov(O_{t_1},O_{t_2}) = \Cov(O_{t_2},O_{t_1})$, we assume without loss of generality that $t_1 \leq t_2$ and note that the conditional probabilities in \eqref{eq:cov_joint_o} can be computed by evaluating \eqref{eq:cme} for different deterministic initial values $(n_1,o_1)$.
This means, we obtain $\Pr\lbrace\left. (N(t_2),O(t_2)) = (n_2,o_2) \,\right| (N(t_1),O(t_1)) = (n_1,o_1) \rbrace$ by setting $t_0=t_1$ in \eqref{eq:cme}, choosing the following initial value for \eqref{eq:cme}
\begin{align}
    \P(n, o, t_0) = \begin{cases}
        1,\qquad&\textrm{if}\,(n,o) = (n_1,o_1),\\
        0,\qquad&\textrm{otherwise},
    \end{cases}\label{eq:cov:init_val}
\end{align}
and evaluating the solution of \labelcref{eq:cme,eq:cov:init_val} at $t=t_2-t_1$.
Hence, $\Cov(O_{t_1},O_{t_2})$ is obtained by repeatedly applying Algorithm~\ref{alg:cme_lumped_state_space} for different initial values $\P(n, o, t_0)$.
The steps required for computing $\Cov(O_{t_1},O_{t_2})$ are summarized in Algorithm~\ref{alg:koo}.

\begin{algorithm}[t]
    \small
    \caption{Computation of $\Cov(O_{t_1},O_{t_2})$}
    \begin{algorithmic}[1]\label{alg:koo}
        \STATE \textbf{input:} $t_1$, $t_2$, $\epsilon$
        \STATE Compute $\Pr\lbrace \left. (N(t_1),O(t_1)) = (n,o)\,\right| (N(t_0),O(t_0)) = (n_0,o_0) \rbrace$ using Algorithm~\ref{alg:cme_lumped_state_space} with $t_0=0$ and $\pibd_0=[1,0,\ldots,0]^{\mathrm{T}}$.
        \FORALL{$(n_1,o_1)$ such that $\P(n_1, o_1, t_1) > \epsilon$}
        \STATE Compute $\Pr\lbrace\left. (N(t_2),O(t_2)) = (n,o) \,\right| (N(t_1),O(t_1)) = (n_1,o_1) \rbrace$ using Algorithm~\ref{alg:cme_lumped_state_space} with $t_0=t_1$ and $\pibd_0$ as defined in \eqref{eq:cov:init_val}.
        \ENDFOR
        \STATE Compute $P_O(o_1,o_2,t_1,t_2)$ using \eqref{eq:cov_joint_o}.
        \STATE Compute $\Cov(O_{t_1},O_{t_2})$ using \eqref{eq:def:cov}.
        \RETURN $\Cov(O_{t_1},O_{t_2})$.
    \end{algorithmic}
\end{algorithm}

\section{Numerical Results}
\label{sec:numerical_results}
\subsection{Particle-Based Simulation and Choice of Parameters}

Stochastic \acp{PBS} are conducted to simulate random trajectories of the \ac{PSP}.
To this end, the three-dimensional Brownian motion of \acp{NT} in the synaptic cleft, the reversible binding of \acp{NT} to individual postsynaptic receptors, and the random degradation of solute \acp{NT} are simulated according to the simulator design outlined in \cite{lotter20,lotter21}.
The random realizations of the synaptic reaction-diffusion process obtained via \acp{PBS} are then used to compute the \ac{PSP} by applying the nonlinear filter \eqref{eq:psp}.

For the \acp{PBS}, we consider three sets of parameter values, $\mathfrak{S}_0$, $\mathfrak{S}_1$, and $\mathfrak{S}_2$, listed in Table~\ref{tab:parameter_values}.
Further model parameters relevant for the \ac{PBS} and the state space model $\mathcal{S}$ but not for the \ac{CME} model considered in this paper are set according to \cite[Table~1]{lotter21}.

\begin{table}
    \vspace*{0.07in}
    \centering
    \caption{\Ac{PBS} parameter values for scenarios considered in Sec.~\ref{sec:numerical_results}.}
    \vspace*{-1mm}
    \footnotesize
    \begin{tabular}{| p{.18\linewidth} | r | r | r |}
        \hline & $\mathfrak{S}_0$\cite{lotter21,savtchenko13} & $\mathfrak{S}_1$ & $\mathfrak{S}_2$\\ \hline
        $N_0$~$[-]$ & $\SI{2000}{}$ & $\SI{1000}{}$ & $\SI{250}{}$\\ \hline
        $C$~$[-]$ & $203$ & $600$ & $600$\\ \hline
        $\ka$~$[\si{\micro\meter\per\micro\second}]$ & $1.52 \times 10^{-5}$ & $4.48 \times 10^{-3}$ & $4.48 \times 10^{-4}$\\ \hline
        $\kd$~$[\si{\per\micro\second}]$ & $8.5 \times 10^{-3}$ & $8.5 \times 10^{-3}$ & $8.5 \times 10^{-3}$ \\ \hline
        $\ke$~$[\si{\per\micro\second}]$ & $10^{-3}$ & $10^{-3}$ & $10^{-5}$ \\ \hline
    \end{tabular}
    \label{tab:parameter_values}
\end{table}

$\mathfrak{S}_1$ is used to model synapses in which the competition of \acp{NT} for receptors is relatively small as compared to $\mathfrak{S}_0$ as it is the case in the neuromuscular junction where more receptors are present than in central synapses \cite{holmes95}.
In addition to the ratio of receptors to released \acp{NT}, also the binding rate $\ka$ is increased in $\mathfrak{S}_1$ as compared to $\mathfrak{S}_0$ to account for the presence of ionotropic high-affinity receptors in some synapses \cite{dincamps14}.
In the parameter regime of $\mathfrak{S}_1$, the assumption underlying the model proposed in \cite{lotter21} is not fulfilled.

$\mathfrak{S}_2$ models a scenario in which many receptors compete for relatively few \acp{NT}.
Although \acp{NT} are usually more abundant than receptors, this situation may occur as a consequence of impaired vesicle loading \cite{pothos02}.
It is assumed that \acp{NT} are degraded relatively slowly in $\mathfrak{S}_2$ and that the receptors employed in $\mathfrak{S}_2$ possess medium affinity for the released \acp{NT} compared to $\mathfrak{S}_0$ and $\mathfrak{S}_1$.
These assumptions reflect the natural variability of biological synapses and the results presented later in this section show that the accuracy of the model proposed in this paper is not affected by the changes in the corresponding parameter values.

For any realization of the random receptor occupancy after the release of \acp{NT}, $O(t)$, the output of the nonlinear filter \eqref{eq:psp} is computed using the analytical solution of \eqref{eq:psp} given as
\begin{multline}
    \Vm(t) =\exp\left(-\int_{0}^{t} \alpha - \gamma O(\tau) \dtau\right)\\
    \times \int_{0}^{t} \exp\left(\int_{0}^{\tau} \alpha -\gamma O(\theta) \mathrm{d}\theta\right) \gamma O(\tau) (\EL - \Er) \dtau + \EL,
\end{multline}
where $\alpha$ was defined in Section~\ref{sec:psp}, $\gamma = -\gsc/(\A\Cm)$, and $O(t)$ is obtained by \acp{PBS}.
The default parameter values for the computation of \eqref{eq:psp} are listed in Table~\ref{tab:parameter_values_psp}.

\begin{table}
    \centering
    \caption{Parameter values for computation of \ac{PSP} and Alg.~\ref{alg:cme_lumped_state_space} proposed in Sec.~\ref{sec:cme}.}
    \footnotesize
    \begin{tabular}{| p{.2\linewidth} | r | p{.38\linewidth} |}
        \hline Parameter & Default Value & Description \\ \hline
        $\Cm$~$[\si{\pico\farad\per\square\micro\meter}]$ & $10^{-2}$ \cite{destexhe01} & Membrane capacitance\\ \hline
        $\EL$~$[\si{\milli\volt}]$ & $-80$ \cite{destexhe01} & Resting potential \\ \hline
        $\Er$~$[\si{\milli\volt}]$ & $0$ \cite{destexhe01} & Reversal potential \\ \hline
        $\gL$~$[\si{\nano\siemens\per\micro\meter\squared}]$ & $4.53 \times 10^{-4}$ \cite{destexhe01} & Leakage conductance \\ \hline
        $\gsc$~$[\si{\nano\siemens}]$ & $0.1$ \cite{byrne14} & Single channel conductance \\ \hline
        $\A$~$[\si{\micro\meter\squared}]$ & $50,000$ \cite{destexhe01} & Postsynaptic membrane surface area \\ \hline
        $\epsilon$~$[-]$ & $5\times10^{-11}$ & Threshold in Algorithm~\ref{alg:cme_lumped_state_space}\\ \hline
        $\Delta t$~$[\si{\micro\second}]$ & $50$ & Time step for Algorithm~\ref{alg:cme_lumped_state_space}\\ \hline
    \end{tabular}
    \label{tab:parameter_values_psp}
\end{table}

\subsection{Validation of the CME Model}

In this subsection, we present numerical results for the statistics of $N(t)$ and $O(t)$ obtained with Algorithm~\ref{alg:cme_lumped_state_space} and compare these results with two reference models for $O(t)$, one reference model for $N(t)$, and stochastic \ac{PBS}.
The reference models for $O(t)$ are the statistical model based on the hypergeometric distribution proposed in \cite{lotter21}, denoted by $\mathfrak{H}(o)$, and the binomial model obtained by assuming statistical independence of the receptors, $\mathfrak{B}(o) = P_B(o;C,o(t)/C)$, where $P_B$ was defined in Section~\ref{sec:cme:state_reduction}.
In lack of any existing reference model for $N(t)$, we compare the predictions of our model for $P_N(n)$ with the binomial model obtained under the assumption that \acp{NT} are degraded independently of each other, i.e., $\mathfrak{N}(n) = P_B(n;N_0,n(t)/N_0)$.
To validate the results obtained with Algorithm~\ref{alg:cme_lumped_state_space} by \acp{PBS}, we compute the empirical distribution of $N(t)$ and $O(t)$ at given time $t$ based on $6,000$ \ac{PBS} realizations.

\subsubsection{Receptor Occupancy Statistics}

\begin{figure*}
    \centering
    \includegraphics[width=\linewidth]{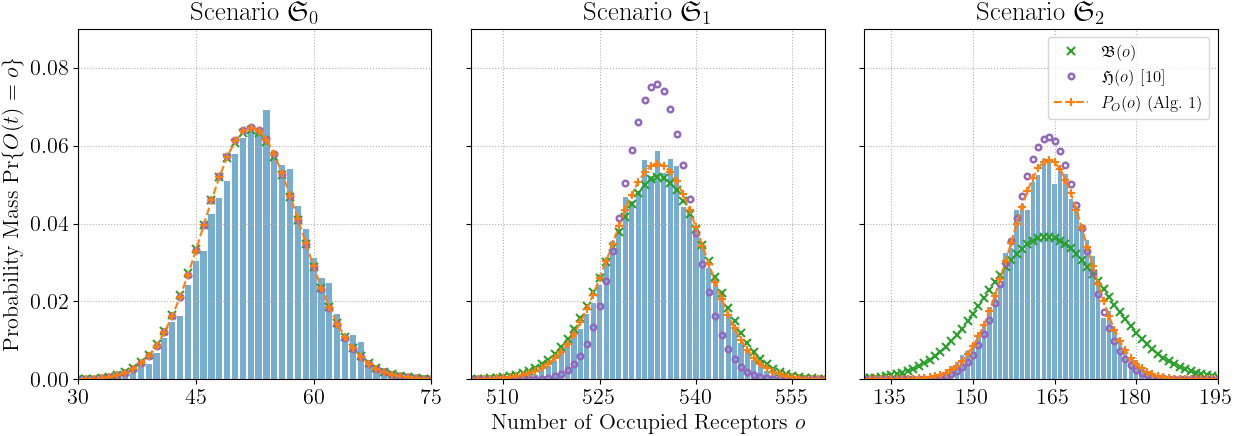}
    \caption{Probability mass function of $O(t)$ at $t=1$~$\si{\milli\second}$ as predicted by the model proposed in Sections~\ref{sec:system_model} and \ref{sec:cme} (orange), the statistical model proposed in \cite{lotter21} (purple), and the binomial model (green). Results from \acp{PBS} are shown in blue. The three subfigures correspond to scenarios $\mathfrak{S}_0$, $\mathfrak{S}_1$, and $\mathfrak{S}_2$, respectively, as defined in Table~\ref{tab:parameter_values}. The orange, purple, and green curves coincide in the left subfigure.}
    \label{fig:rec_occ_stats}
\end{figure*}

Fig.~\ref{fig:rec_occ_stats} shows $\mathrm{P}_O(t)$ at $t=\SI{1}{\milli\second}$ as obtained by Algorithm~\ref{alg:cme_lumped_state_space} and the reference models $\mathfrak{H}$ and $\mathfrak{B}$, as well as the results obtained with \acp{PBS}, for $\mathfrak{S}_0$, $\mathfrak{S}_1$, and $\mathfrak{S}_2$.
We observe from Fig.~\ref{fig:rec_occ_stats} that the model proposed in Sections~\ref{sec:system_model} and \ref{sec:cme} matches the empirical distribution obtained by \ac{PBS} accurately for all considered sets of parameters.
Also, both reference models $\mathfrak{H}$ and $\mathfrak{B}$ match the \ac{PBS} data for $\mathfrak{S}_0$.
However, $\mathfrak{H}$ fails to reproduce $\mathrm{P}_O(t)$ for $\mathfrak{S}_1$.
The reason for this is that due to the abundance of both \acp{NT} and receptors in $\mathfrak{S}_1$, there is neither competition among \acp{NT} for receptors nor competition among receptors for \acp{NT} and the main assumption for $\mathfrak{H}$ is not fulfilled.
On the other hand, $\mathfrak{B}$ fails to reproduce $\mathrm{P}_O(t)$ for $\mathfrak{S}_2$, the reason being that for $\mathfrak{S}_2$, the independence assumption underlying $\mathfrak{B}$ is not fulfilled.
We conclude that the statistical model for $O(t)$ proposed in this paper is more robust with respect to parameter variations than previous models.

Finally, we observe from Fig.~\ref{fig:rec_occ_stats} that the variance of $O(t)$ and, consequently, the statistical dependence between the activation of different postsynaptic receptors depends largely on the choice of the synaptic parameters.
While correlation between receptors is rather strong in $\mathfrak{S}_2$, it is almost negligible in $\mathfrak{S}_0$ and $\mathfrak{S}_1$.
In $\mathfrak{S}_0$, on the other hand, competition among \acp{NT} for receptors is \mbox{stronger compared to $\mathfrak{S}_1$.}

\subsubsection{NT Degradation Statistics}

\begin{figure}[t]
    \includegraphics[width=\linewidth]{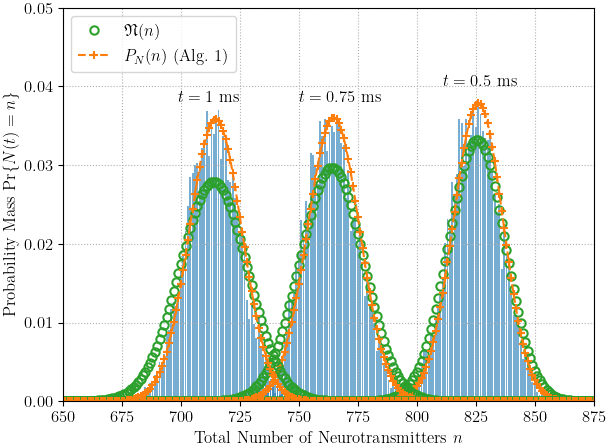}
    \caption{$P_N(t)$ at different time instants $t$ for $\mathfrak{S}_1$. The figure shows the \ac{CME} model proposed in this paper (orange), the binomial model $\mathfrak{N}$ (green), and results from \acp{PBS} (blue).}
    \label{fig:inst_deg_stats}
\end{figure}

Next, we consider the statistics of $N(t)$.
Fig.~\ref{fig:inst_deg_stats} shows the marginal distribution of $N(t)$ as obtained by Algorithm~\ref{alg:cme_lumped_state_space}, reference model $\mathfrak{N}$, and \ac{PBS} data at different time instants $t=\SI{0.5}{\milli\second},\SI{0.75}{\milli\second}$, and $\SI{1}{\milli\second}$ for parameter values $\mathfrak{S}_1$.
First, we observe from Fig.~\ref{fig:inst_deg_stats} that the results obtained with Algorithm~\ref{alg:cme_lumped_state_space} match the empirical distribution of $N(t)$ very well for all considered time instants.
Furthermore, we observe from Fig.~\ref{fig:inst_deg_stats} that the degradation of \acp{NT} is negatively correlated, since $P_N(t)$ as obtained with Algorithm~\ref{alg:cme_lumped_state_space} is more concentrated compared to the binomial model $\mathfrak{N}$.
The negative correlation is expected here, since \acp{NT} are only exposed to degradation if they are solute.
On the other hand, as more \acp{NT} are degraded, it is more likely that the remaining \acp{NT} bind to receptors - which in turn prevents them from being degraded.

From these results, we conclude that the proposed model can be used to gain novel insights into the impact of the various synaptic parameters on the statistics of synaptic signaling.

\begin{figure}[t]
    \includegraphics[width=\linewidth]{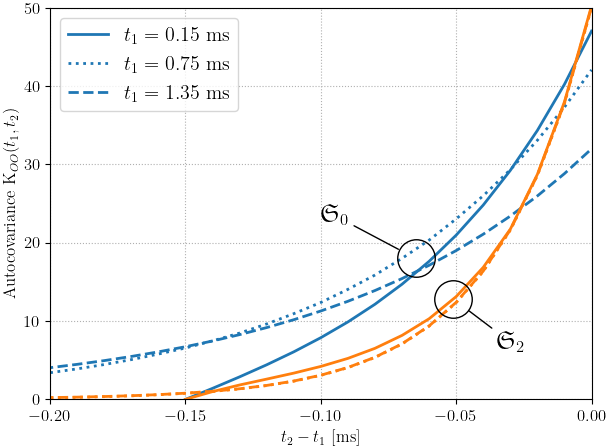}
    \caption{Receptor occupancy autocovariance $\KOO(t_1, t_2)$ as defined in \eqref{eq:def:cov} as a function of $t_2-t_1$ for different time instants $t_1$ in scenarios $\mathfrak{S}_0$ (blue) and $\mathfrak{S}_2$ (orange).}
    \label{fig:rec_occ_cov}
\end{figure}

\subsection{Receptor Occupancy Autocovariance}

In this section, we study the autocovariance of the receptor occupancy $\KOO(t_1,t_2)$ as defined in \eqref{eq:def:cov}.
Fig.~\ref{fig:rec_occ_cov} shows $\KOO(t_1,t_2)$ computed by applying Algorithm~\ref{alg:koo} as described in Section~\ref{sec:cme:cov} for scenarios $\mathfrak{S}_0$ and $\mathfrak{S}_2$.
We observe from Fig.~\ref{fig:rec_occ_cov} that the occupancy of receptors at some time $t_1$ is positively correlated with the occupancy of the receptors at previous time instants $t_2 < t_1$.
This positive correlation can be attributed to the following two reasons.
First, the binding and unbinding of \acp{NT} to and from postsynaptic receptors, respectively, is non-instantaneous and, hence, it is likely that some of the receptors which are bound (unbound) at $t_2$ are still bound (unbound) at $t_1$ if $|t_2 - t_1|$ is sufficiently small.
Second, the buffering of \acp{NT} at postsynaptic receptors prevents these \acp{NT} from being degraded by enzymes.
Now, if the number of buffered (bound) \acp{NT} at $t_2$ is large, the concentration of \acp{NT} applied to postsynaptic receptors at $t_1$ and, consequently, also the number of occupied receptors at $t_1$ is large.
As the time interval between $t_1$ and $t_2$ increases, the receptor occupancies at $t_1$ and $t_2$, respectively, become less and less correlated.

From Fig.~\ref{fig:rec_occ_cov}, we observe furthermore that $\KOO(t_1,t_2)$ as a function of $t_1$ and $|t_2 - t_1|$ decays differently in the two scenarios $\mathfrak{S}_0$ and $\mathfrak{S}_2$.
There are two insights to gain here and both are related to the relatively weak enzymatic degradation in $\mathfrak{S}_2$ as compared to $\mathfrak{S}_0$.
First, $\KOO(t_1,t_2)$ varies as a function of $t_1$ if $|t_2 - t_1|$ is kept fixed for $\S_0$.
In particular, for the parameter values of $\mathfrak{S}_0$, $\KOO(t_1,t_2)$ decreases more slowly as $t_1$ increases.
However, $\KOO(t_1,t_2)$ is (almost) constant in $t_1$ for $\S_2$.
This is a consequence of the fact that the concentration of \acp{NT} applied to postsynaptic receptors in $\S_2$ is almost constant over time while it varies significantly in $\S_0$.
Second, $\KOO(t_1,t_2)$ is larger for $\S_0$ compared to $\S_2$ if $|t_2 - t_1| > \SI{0.04}{\milli\second}$.
This results from the fact that from the two sources of positive correlation mentioned above, namely non-instantaneous ligand-receptor binding and \ac{NT} buffering, only the first one is relevant for $\S_2$ while both contribute to the autocovariance in $\S_0$.
In other words, the receptor occupancy in $\S_0$ is subject to one additional source of randomness compared to $\S_2$, since the enzymatic degradation is not significant in $\S_2$ at the time scale considered here.

The above analysis of $\KOO$ provides a more comprehensive statistical characterization of the molecular signaling process in synaptic \ac{DMC} compared to the instantaneous statistics of $O(t)$.
Indeed, the results presented in this paper underline the importance of such a comprehensive model for understanding the statistical properties of the downstream signal which we believe will eventually be important for the design of synthetic \ac{MC} systems based on ligand-binding receptor-based receivers.

\vspace{-4mm}
\subsection{The Postsynaptic Membrane Potential}

In this section, we study the statistics of the \ac{PSP} as predicted by the \ac{CME} model proposed in Section \ref{sec:cme} and the statistical model for the \ac{PSP} proposed in Section~\ref{sec:psp}, and compare the model predictions to results from \acp{PBS}.

\subsubsection{Statistics of the PSP}

\begin{figure}[t]
    \includegraphics[width=\linewidth]{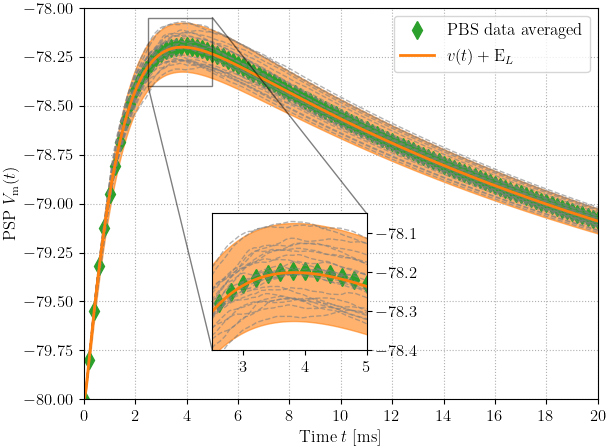}
    \caption{\ac{PSP} as predicted by the model proposed in Section~\ref{sec:cme} (orange) and obtained from stochastic \acp{PBS}, respectively. The shaded orange region corresponds to the area between $v(t) + \EL - 2\sigma_V(t)$ and $v(t) + \EL + 2\sigma_V(t)$, cf.~\labelcref{eq:psp:mean,eq:psp:var}. The ensemble average and individual realizations of the \acp{PBS} are shown as green diamond markers and gray dashed lines, respectively.}
    \label{fig:psp}
\end{figure}

In this section, we consider synaptic transmission according to $\mathfrak{S}_0$ and assume that the membrane of the postsynaptic neuron is configured according to the default parameter values listed in Table~\ref{tab:parameter_values_psp}.
Fig.~\ref{fig:psp} shows the expected \ac{PSP} after the release of \acp{NT} at $t=0$ as predicted by \eqref{eq:psp:mean} and the ensemble average of the random \ac{PSP} trajectories obtained from the \acp{PBS}.
Furthermore, Fig.~\ref{fig:psp} shows some individual random \ac{PSP} realizations and an error margin of $\pm 2\sigma_V(t)$ around its predicted mean value accounting for the non-stationary randomness of the \ac{PSP}.
If the \ac{PSP} was Gaussian distributed with mean $v(t) + \EL$ and variance $\sigma^2_V(t)$, approximately $95\%$ of the random \ac{PSP} realizations would lie within this error margin as the number of realizations tends to infinity.

First, we observe from Fig.~\ref{fig:psp} that the proposed model matches the simulated \ac{PSP} trace very accurately.
This observation justifies the use of the linear \ac{PSP} model \eqref{eq:psp_linear}.
Furthermore, we observe from Fig.~\ref{fig:psp} that the applied error margin based on the Gaussian approximation and \eqref{eq:psp:var} provides an accurate approximation for the stochastic variability of the \ac{PSP}.
We observe from Fig.~\ref{fig:psp} that the \ac{PSP} is characterized by an initial rising phase of approximately \SI{3.5}{\milli\second}, during which the postsynaptic membrane is depolarized by approximately $1.8 \pm 0.1$~\si{\milli\volt}, and a subsequent slow decay phase, during which the membrane is re-polarized.
These characteristics are in good agreement with values reported in the literature \cite{byrne14}.

We also observe from Fig.~\ref{fig:psp} that the randomness of the \ac{PSP} at some time instant $t_1$ does not only depend on the expected value of the signal $\Vm(t_1)$, but also on the value of $t_1$ itself.
Consider for example $t_1=1.5 \si{\milli\second}$ and $t_2=9 \si{\milli\second}$.
Then, $\Vm(t_1) \approx \Vm(t_2) \approx \SI{-78.5}{\milli\volt}$, but the variance of $\Vm(t_1)$ is much smaller than the variance of $\Vm(t_2)$ as can be seen from both the random realizations of $\Vm(t)$ and the statistical model based on \labelcref{eq:psp:mean,eq:psp:var}.
This observation shows the impact of the statistics of the reaction-diffusion process underlying the \ac{PSP} on the variability of the \ac{PSP}.

\begin{figure}[t]
    \includegraphics[width=\linewidth]{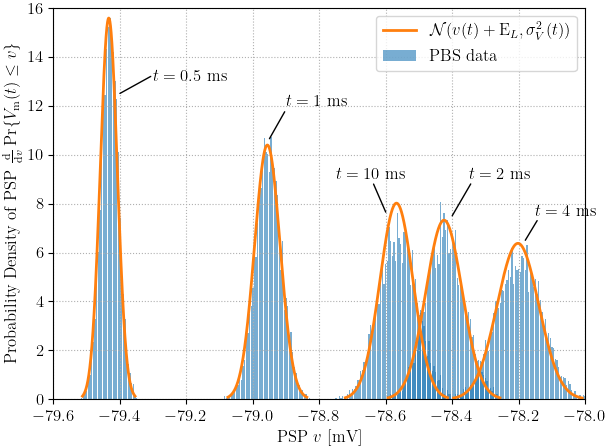}
    \caption{Instantaneous statistics of the \ac{PSP} as predicted by the Gaussian model based on \labelcref{eq:psp:mean,eq:psp:var} (orange) and \ac{PBS} data (blue), respectively.}
    \label{fig:psp_stats}
\end{figure}

Fig.~\ref{fig:psp_stats} shows the instantaneous statistics of the \ac{PSP} as predicted by the Gaussian model proposed in Section~\ref{sec:psp} \eqref{eq:psp:gaussian_approx} and computed from the random \ac{PBS} trajectories, respectively.
We observe from Fig.~\ref{fig:psp_stats} that \eqref{eq:psp:gaussian_approx} provides an accurate model for the statistics of the \ac{PSP}.
In particular, the accuracy of the proposed approximation is very good for the rising phase and the peak value of the \ac{PSP} and only decreases slightly as the membrane potential tends back to its resting value.
This is a consequence of the fact that the \ac{CME} model proposed in Section~\ref{sec:cme} is based on the simplifying assumption that the fraction of \acp{NT} close to the postsynaptic membrane is deterministic, cf.~\eqref{eq:diff_approx}.
This assumption is accurate as long as the number of \acp{NT} is large enough, but it becomes less accurate for large times $t$ as the number of \acp{NT} decreases and the variability of $S_a(t)/S(t)$ in \eqref{eq:diff_approx} increases.

In summary, the proposed model reveals by how much the \ac{PSP} varies due to the randomness of the reaction-diffusion process governing synaptic transmission.
Hence, the proposed model presents a step towards elucidating the contributions of different sources of randomness to the random fluctuations of the \ac{PSP} observed in experimental data, which is a research gap left open by current computational models of synaptic transmission \cite{baker11}.
Furthermore, since the randomness of synaptic transmission is assumed to encode the reliability of the transmitted information \cite{aitchison21}, the proposed model contributes to the understanding of the role of the molecular noise in synaptic \ac{DMC} for neuronal information transmission and processing.
In particular and in contrast to existing models, it allows for a {\em quantitative assessment} of the impact of this noise on the \ac{PSP}.

\subsubsection{Statistics of the PSP for a ``fast'' Neuron}\label{sec:numerical_results:psp:fast}
\begin{figure}[t]
    \includegraphics[width=\linewidth]{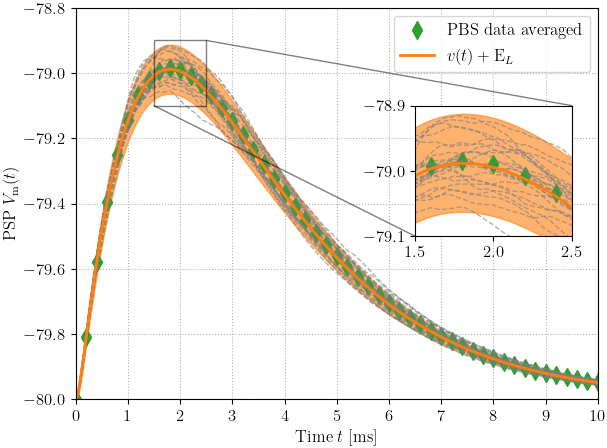}
    \caption{\ac{PSP} as predicted by the model proposed in Section~\ref{sec:cme} (orange) and obtained from stochastic \acp{PBS} for $\gL = 4.53 \times 10^{-3}\,\si{\nano\siemens\per\micro\meter\squared}$, respectively. The shaded orange region corresponds to the area between $v(t) + \EL -2\sigma_V(t)$ and $v(t) + \EL + 2\sigma_V(t)$, cf.~\labelcref{eq:psp:mean,eq:psp:var}. The ensemble average and individual realizations of the \acp{PBS} are shown as green diamond markers and gray dashed lines, respectively.}
    \label{fig:psp_fast}
\end{figure}

In this section, we consider a neuron with an increased membrane leakage conductance of $\gL = 4.53 \times 10^{-3}$~\si{\nano\siemens\per\micro\meter\squared}.
Synaptic transmission is assumed to occur according to $\mathfrak{S}_0$.
Fig.~\ref{fig:psp_fast} shows the expected \ac{PSP} as predicted by \eqref{eq:psp:mean} and the \ac{CME} model presented in Section~\ref{sec:cme} and by the ensemble average of the \acp{PBS}, respectively.
Furthermore, some individual random realizations of the \ac{PSP} and an error margin of $\pm 2\sigma_V(t)$, cf.~\eqref{eq:psp:var}, around the predicted mean value are shown.
We observe from Fig.~\ref{fig:psp_fast} that the proposed model matches the ensemble average of the \ac{PBS}-based \ac{PSP} realizations very accurately.
Furthermore, we observe that almost all random trajectories of the \ac{PSP} fall within the error margin.
Finally, we observe from Fig.~\ref{fig:psp_fast} that the \ac{PSP} decays faster and the peak value is lower compared to the reference scenario shown in Fig.~\ref{fig:psp}.
This is a consequence of the fact that the synaptic current resulting from the activation of postsynaptic receptors leaks more rapidly through the membrane of the postsynaptic neuron as $\gL$ is increased, and therefore the temporary depolarization of the postsynaptic neuron lasts for a shorter amount of time as compared to the default case considered in the previous subsection.

\begin{figure}[t]
    \includegraphics[width=\linewidth]{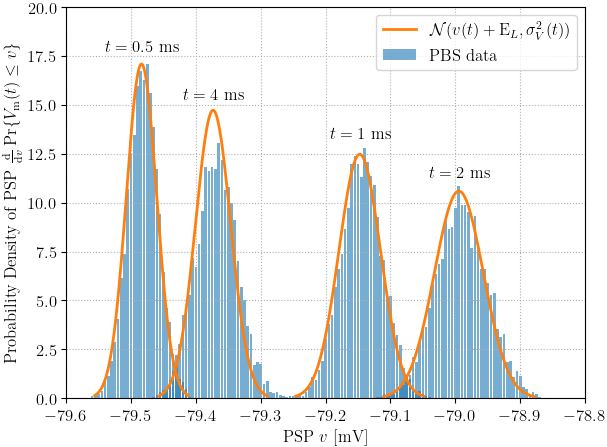}
    \caption{Instantaneous statistics of the \ac{PSP} as predicted by the Gaussian model based on \labelcref{eq:psp:mean,eq:psp:var} (orange) and \ac{PBS} data (blue), respectively, for $\gL = 4.53 \times 10^{-3}\,\si{\nano\siemens\per\micro\meter\squared}$.}
    \label{fig:psp_stats_fast}
\end{figure}

Fig.~\ref{fig:psp_stats_fast} shows the instantaneous statistics of $\Vm(t)$ at different time instants as predicted by the Gaussian model \eqref{eq:psp:gaussian_approx} and the \ac{PBS} data, respectively.
We observe from Fig.~\ref{fig:psp_stats_fast} that \eqref{eq:psp:gaussian_approx} provides an accurate estimate of the \ac{PSP} statistics.
Furthermore, we observe from Fig.~\ref{fig:psp_stats_fast} that at all time instants $t=0.5,1,2,4$~\si{\milli\second}, the spread of $\Vm(t)$ is relatively large as compared to the same time instants in Fig.~\ref{fig:psp_stats}.
This observation shows that the stochastic variability of the \ac{PSP} does not only depend on the randomness of the reaction-diffusion process in the synaptic cleft (which is identical for Figs.~\ref{fig:psp_stats} and \ref{fig:psp_stats_fast}), but also on the electrophysiological properties of the postsynaptic membrane.

\begin{figure}[!t]
    \centering
    \includegraphics[width=\linewidth]{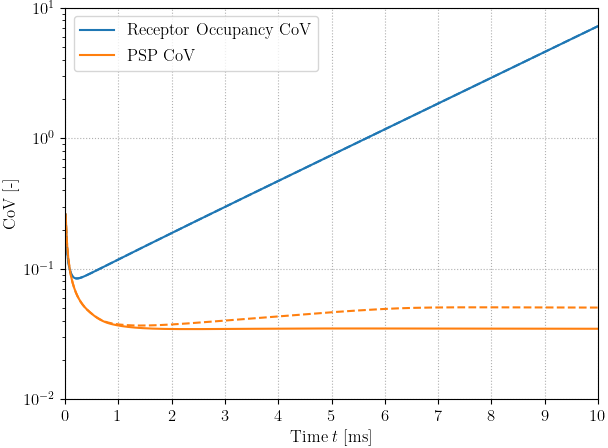}
    \caption{Coefficient of variation of the receptor occupancy $O(t)$ (blue) and the depolarization of the postsynaptic membrane $V(t)$ (orange), respectively, in $\mathfrak{S}_0$. The solid orange line corresponds to the default membrane parameter values listed in Table~\ref{tab:parameter_values_psp}, the dashed orange line corresponds to an increased leakage conductance $\gL = 4.53 \times 10^{-3}\,\si{\nano\siemens\per\micro\meter\squared}$ as considered in Subsection~\ref{sec:numerical_results:psp:fast}.}
    \label{fig:cov}
\end{figure}

Fig.~\ref{fig:cov} shows the {\em \ac{CoV}}, i.e., the ratio of the standard deviation to the mean, of the receptor occupancy $O(t)$ and of the depolarization of the postsynaptic membrane $V(t)$ as defined in Section~\ref{sec:psp:linear_approximation}, respectively.
We observe from Fig.~\ref{fig:cov} that the low-pass property of the postsynaptic membrane reduces the stochastic variability of the \ac{PSP} as compared to $O(t)$.
In fact, we observe from Fig.~\ref{fig:cov} that the \ac{CoV} of $O(t)$ diverges, while the \ac{CoV} of the \ac{PSP} tends towards a constant value for both the default membrane parameters (solid orange line) and the increased leakage conductance (dashed orange line) considered in this section.
These observations indicate that the processing of the molecular synaptic signal by the postsynaptic membrane leads to an electrochemical downstream signal with significantly reduced stochastic variability as compared to the chemical signal inside the synaptic cleft.


\subsubsection{The Impact of Synaptic Configuration and Membrane Properties on the PSP}

\begin{figure*}
    \centering
    \includegraphics[width=\textwidth]{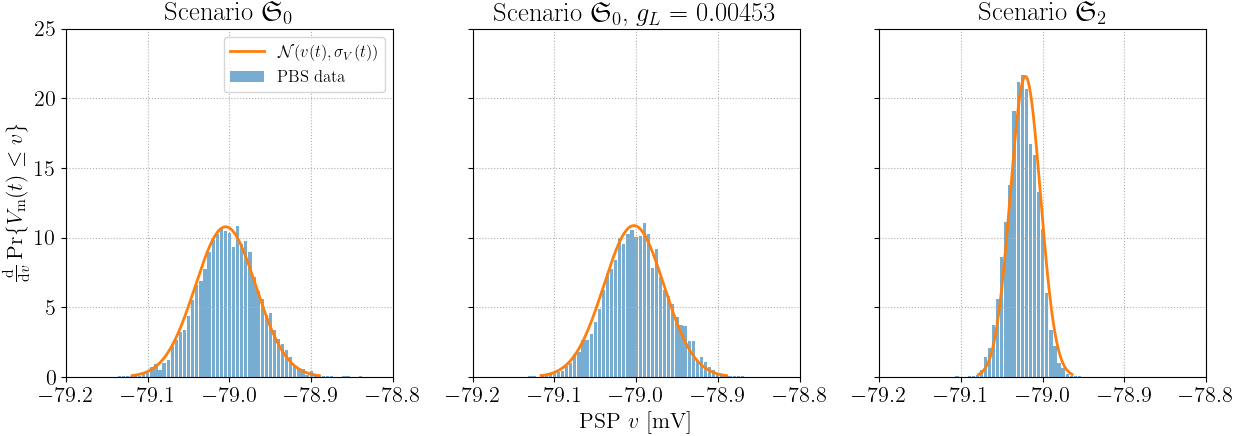}
    \caption{Instantaneous statistics of the \ac{PSP} as obtained by \eqref{eq:psp:gaussian_approx} (orange) and \acp{PBS} (blue), respectively, when $\Vm(t) \approx -79.0\,\si{\milli\volt}$ for different configurations of the synapse and the postsynaptic membrane. The data in the left and middle panel is computed using the synaptic configuration of Scenario $\S_0$, the right panel is based upon $\S_2$. The left and the right panel assume default values for the electrophysiological properties of the postsynaptic membrane, the middle panel assumes a leakage conductance of $\gL = 4.53 \times 10^{-3}\,\si{\nano\siemens\per\micro\meter\squared}$.}
    \label{fig:psp_stats_v0}
\end{figure*}

In this section, we study how much the randomness of the postsynaptic receptor activation contributes to the randomness of the \ac{PSP} as compared to the filtering by the postsynaptic membrane.
To this end, we compare the statistics of $\Vm(t)$ for Scenario $\S_0$ with default parameter values for the postsynaptic membrane, Scenario $\S_0$ with $\gL = 4.53 \times 10^{-3}$~\si{\nano\siemens\per\micro\meter\squared}, and Scenario $\S_2$, respectively.
Fig.~\ref{fig:psp_stats_v0} shows the statistics of $\Vm(t)$ as predicted by \eqref{eq:psp:gaussian_approx} and \ac{PBS}, respectively, when $\Vm(t)$ assumes a value of approximately $-79.0$~\si{\milli\volt}.
We observe from Fig.~\ref{fig:psp_stats_v0} that the statistics of $\Vm(t)$ are almost identical for Scenario $\S_0$ and Scenario $\S_0$ with $\gL = 4.53 \times 10^{-3}$~\si{\nano\siemens\per\micro\meter\squared}.
In contrast, the statistics of $\Vm(t)$ for Scenario $\S_2$ are much more concentrated than in the other two cases.
This is indeed expected since the binding of \acp{NT} to postsynaptic receptors is much more deterministic for $\S_2$ than for $\S_0$, and the autocovariance (which contributes positively to the variance of $\Vm(t)$) decays much faster for $\S_2$ than for $\S_0$, cf.~Fig.~\ref{fig:rec_occ_cov}.
This observation indicates that the random activation of postsynaptic receptors plays a vital role for the variability of the \ac{PSP}.
Furthermore, it shows that, despite the filtering of the synaptic signal by the postsynaptic cell, the statistics of the \ac{PSP} depend largely on the configuration of the synapse.
Since the randomness of the \ac{PSP} is assumed to carry information \cite{aitchison21} and noise in synaptic signaling appears to contribute to the detection of subthreshold signals in some synapses \cite{stacey01}, this observation is an important step towards revealing the significance of the synaptic reaction-diffusion process for the synaptic information transmission.

\section{Conclusion}
\label{sec:conclusion}
In this paper, we proposed a novel statistical model for the \ac{PSP} after neurotransmission.
The proposed model is based on a \ac{CME} model for the receptor occupancy and the \ac{NT} degradation in the synaptic \ac{DMC} system.
The first two moments of the \ac{PSP} were approximated using the time-dependent mean and autocovariance of the postsynaptic receptor occupancy and an approximation of the \ac{PSP} in terms of the Gaussian distribution was derived.
Since the \ac{CME} model cannot be solved in closed-form and is computationally intractable, an adaptive state reduction scheme was proposed to compute the first- and second-order moments of the postsynaptic receptor occupancy.
The proposed Gaussian approximation of the \ac{PSP} was validated with \acp{PBS} and shown to provide very accurate results.

The model proposed in this paper is the first one to explicitly link biophysical parameters of the synaptic \ac{DMC} system to the statistics of the \ac{PSP}.
The results presented in this paper show that different synaptic configurations impact the autocovariance of the postsynaptic receptor activation and hereby shape the statistics of the \ac{PSP}.
Furthermore, the proposed model reveals that due to the postsynaptic processing of the synaptic signal, the \ac{CoV} of the \ac{PSP} is small compared to the \ac{CoV} of the receptor occupancy, while the \ac{PSP} statistics remain characteristic of the respective synaptic configurations.
As the statistics of the \ac{PSP} ultimately determine the information processing and firing of postsynaptic neurons, it is important to understand how the design of the synaptic \ac{DMC} system contributes to the shaping of these statistics and the proposed model provides a novel tool to study this question.

The proposed model may also prove useful for the design of artificial synapses, since it allows for a better understanding of how different synaptic parameters manifest themselves in the postsynaptic signal.
In this way, the artificial synapse can be tuned to fit the target application best or a detector based on the postsynaptic signal can be designed.
Furthermore, as a communication theoretic tool that establishes a link between the physical parameters of chemical synapses and their statistical properties as communication channels, we believe that the proposed model can contribute to ongoing research efforts in understanding and mitigating synaptopathies \cite{veletic2019,hawk21}.

Since the modeling strategies used in this paper are not specific to synaptic \ac{DMC}, the modeling study presented here may also be helpful for the understanding of the statistical properties of other \ac{DMC} systems, which detect signaling molecules with ligand receptors (see \cite{kuscu19} for examples and a recent review on the physical design of \ac{MC} receivers).

Possible directions for further research include studying the simultaneous activation of multiple synapses and extending the proposed model to the axonal pathway and the \ac{NT} release machinery of presynaptic neurons.

\appendix
\subsection{Proof of Theorem \ref{thm:psp}}\label{sec:app:prf_thm_psp}

To compute $v(t)$ and $\Var(V(t))$, we start from \eqref{eq:psp_sol} and express the stochastic integral in \eqref{eq:psp_sol} as a Riemann sum \cite{schuss09}.
This yields
\begin{equation}
    V(t) = \lim\limits_{\Delta s_i \to 0} \sum_{i} \beta \exp\left[-\alpha(t-s_i)\right] O(s_i) \Delta s_i,\label{eq:riemann_sum}
\end{equation}
where the $s_i$ provide a partition of the interval $[0,t]$.
The limit on the right-hand side of \eqref{eq:riemann_sum} exists, because $O(t)$ has almost surely only a finite number of discontinuities.
Now, taking the expectation of both sides of \eqref{eq:riemann_sum} and then taking the limit of the right-hand side, we obtain due to the linearity of the expectation operator
\begin{equation}
    v(t) = \beta \int_{0}^{t} \exp\left[-\alpha(t-\tau)\right] \Exp{O(\tau)} \dtau.
\end{equation}
Similarly, we compute \eqref{eq:psp_var_proof} on the top of the next page.
\begin{figure*}
    \begin{align}
        \Exp{\left(\sum_{i} \beta \exp\left[-\alpha(t-s_i)\right] O(s_i) \Delta s_i\right)\left(\sum_{j} \beta \exp\left[-\alpha(t-s_j)\right] O(s_j) \Delta s_j\right)}\nonumber\\ = \beta^2 \sum_{i,j} \exp\left[-\alpha(t-s_i)\right] \exp\left[-\alpha(t-s_j)\right] \Exp{O(s_i)O(s_j)} \Delta s_i \Delta s_j\label{eq:psp_var_proof}
    \end{align}    
    \hrulefill
\end{figure*}

Taking the limit and subtracting $\Exp{V(t)}^2$ from \eqref{eq:psp_var_proof}, Theorem~\eqref{thm:psp} follows.
This concludes the proof.

\subsection{Proof of Theorem \ref{thm:state_reduction}}\label{sec:app:prf_thm_state_reduction}

From the structure of $\Abd(t)$, cf.~\eqref{eq:A:Q}, \eqref{eq:A:D}, we know that there is only probability flux from level $n+1$ to level $n$, not vice versa.
Hence, we conclude that
\begin{equation}
    \sum_{n=n_0}^{N} P_N(n,t+\Delta t) \leq \sum_{n=n_0}^{N} P_N(n,t),\label{eq:P_N:upper_tail}
\end{equation}
for any $n_0 \in \lbrace 0,\ldots,N \rbrace$, $\Delta t > 0$.
Eq.~\eqref{eq:P_N:upper_tail} provides an upper tail bound for $P_N(n,t+\Delta t)$ in terms of $P_N(n,t)$.
On the other hand, by the same argument
\begin{equation}
    \sum_{n=0}^{n_0} P_N(n,t) \leq \sum_{n=0}^{n_0} P_N(n,t+\Delta t).\label{eq:P_N:lower_tail}
\end{equation}
Let us consider the interval $I_k$.
By assumption, we know $P_N(n,t_k)$.
Then, with $N^{(k)}_{\mathrm{max}}$ as defined in \eqref{eq:def:Nmax}, we conclude from \eqref{eq:P_N:upper_tail} that $\sum_{n'=\Nmax}^{N} P_N(n',t) < \epsilon$ for any $t \in I_k$.
Now, let us consider the assumption that the \acp{NT} are degraded independently of each other.
Under this assumption, since all \acp{NT} are identical, $N(t)$ follows a binomial distribution with parameters $N_0$ and $\mathbb{E}[N(t)]/N_0 = n(t)/N_0$.
Indeed, this is a worst-case assumption with respect to the spread of $P_N(n,t)$, since in reality, the degradation of \acp{NT} is negatively correlated\footnote{To see the negative dependence of degradation events, consider one \ac{NT} $N_i$. The more \acp{NT} are degraded by time $t$, the more likely it is that $N_i$ binds to a free receptor and thus cannot be degraded by enzymes. On the other hand, the fewer \acp{NT} are degraded, the more \acp{NT} compete for receptors and it is less likely that $N_i$ finds a free receptor that prevents it from being degraded. This conclusion is also confirmed by the results presented in Fig.~\ref{fig:inst_deg_stats}.}, i.e.,
\begin{equation}
    \sum_{n=0}^{n_0} P_N(n,t) \leq \sum_{n=0}^{n_0} P_B(n;N_0,n(t)/N_0),\label{eq:P_N:lower_tail_binom}
\end{equation}
where $P_B(\cdot;n,p)$ as defined in Section~\ref{sec:cme:state_reduction} \cite{yu08}.
Now, with $\Nmin^{(k)}$ as defined in \eqref{eq:def:Nmin}, we conclude from \eqref{eq:P_N:lower_tail_binom} and \eqref{eq:P_N:lower_tail} that $\sum_{n'=0}^{\Nmin^{(k)}} P_N(n',t) < \epsilon$ for any $t \in I_k$.
Since the binding of \acp{NT} to receptors is also negatively correlated \cite{lotter21}, the upper and lower tail bounds for $O(t)$ follow from the same line of argumentation as \eqref{eq:P_N:lower_tail_binom}.
This concludes the proof.

\bibliographystyle{IEEEtran}    
\bibliography{IEEEabrv,H:/libraries/mc}
\end{document}